\documentclass{aa}  
\usepackage{graphicx}
\usepackage{amsmath}	
\usepackage{amssymb}	
\usepackage{relsize}
\usepackage{txfonts}
\newcommand{\msun}{\hbox{M$_{\odot}$}}

\begin{document}

   \title{The universal variability of the stellar initial mass function probed by the TIMER survey}

   \subtitle{}

   \author{
Ignacio Mart\'in-Navarro \inst{1,2}, Adriana de Lorenzo-C\'aceres\inst{1,2}, Dimitri A. Gadotti \inst{3,4}, Jairo Méndez-Abreu\inst{1,2},\\ Jesús Falcón-Barroso \inst{1,2}, Patricia Sánchez-Blázquez\inst{5}, Paula Coelho\inst{6}, Justus Neumann \inst{7}, Glenn van de Ven \inst{8} \\ and Isabel Pérez \inst{9,10}
          }
   \institute{
Instituto de Astrof\'{\i}sica de Canarias, C/ V\'{\i}a L\'actea s/n, E38205 - La Laguna, Tenerife, Spain\\
\email{imartin@iac.es}
\and
Departamento de Astrofísica, Universidad de La Laguna, E-38200 La Laguna, Tenerife, Spain
\and
European Southern Observatory, Karl-Schwarzschild-Str. 2, D-85748 Garching bei München, Germany
\and
Centre for Extragalactic Astronomy, Department of Physics, Durham University, South Road, Durham DH1 3LE, UK 
\and
Departamento de Física de la Tierra y Astrofísica \& IPARCOS, UCM, 28040, Madrid
\and 
Instituto de Astronomia, Geofísica e Ciências Atmosféricas, Universidade de São Paulo, R. do Matão 1226, São Paulo, Brazil
\and
Max-Planck Institut für Astronomie, Königstuhl 17, D-69117, Heidelberg, Germany
\and
Department of Astrophysics, University of Vienna, Turkenschanzstrasse 17, 1180 Wien, Austria
\and
Departamento de Física Teórica y del Cosmos, Campus de Fuente Nueva, Edificio Mecenas, Universidad de Granada, E-18071, Granada, Spain
\and
Instituto Carlos I de Física Teórica y Computacional, Facultad de Ciencias, E-18071 Granada, Spain
}

   \date{Received; accepted}
   
   \titlerunning{The IMF of unresolved young stellar populations}
\authorrunning{Mart\'in-Navarro et al.}  
 
  \abstract
   {The debate about the universality of the stellar initial mass function (IMF) revolves around two competing lines of evidence. While measurements in the Milky Way, an archetypal spiral galaxy, seem to support an invariant IMF, the observed properties of massive early-type galaxies (ETGs) favor an IMF somehow sensitive to the local star formation conditions. The fundamental methodological and physical differences between both approaches have hampered, however, a comprehensive understanding of IMF variations. We describe here an improved modelling scheme that allows for the first time consistent IMF measurements across stellar populations with different ages and complex star formation histories. Making use of the exquisite MUSE optical data from the TIMER survey and powered by the MILES stellar population models, we show the age, metallicity, [Mg/Fe], and IMF slope maps of the inner regions of NGC\,3351, a spiral galaxy with a mass similar to that of the Milky Way. The measured IMF values in NGC\,3351 follow the expectations from a Milky Way-like IMF, although they simultaneously show systematic and spatially coherent variations, particularly for low-mass stars. In addition, our stellar population analysis reveals the presence of metal-poor and Mg-enhanced star-forming regions that appear to be predominantly enriched by the stellar ejecta of core-collapse supernovae. Our findings showcase therefore the potential of detailed studies of young stellar populations to better understand the early stages of galaxy evolution and,  in particular, the origin of the observed IMF variations beyond and within the Milky Way. 
   }


   \maketitle
%

\section{Introduction} \label{sec:intro}
Much has been debated about the universality of the stellar initial mass function (IMF), which in essence describes the (idealized) mass spectrum of stars at birth. The IMF is a central idea in our understanding and modeling of the baryonic cycle in galaxies, bridging the quantum processes driving stellar evolution to the cosmological scales of galaxy formation \citep[see e.g.][for a recent overview]{Kroupa19}. Therefore, measuring the IMF is a critical and necessary endeavour but it has proven to be, observationally, particularly challenging  \citep[e.g.][]{bastian,Smith20}.

The analysis of nearby systems, where individual stars can be resolved and thus {\it direct} measurements of the IMF are feasible, has revealed a rather invariant IMF behavior. This Milky Way-like standard IMF is characterized by a power law distribution with a logarithmic slope of $\Gamma \simeq1.3$ for stars more massive than $\sim0.5$\msun \, and a shallower distribution for lower mass stars \citep[e.g.][]{Miller79,MW,kroupa,Chabrier}. These observational findings cemented the now widely adopted assumption of a universal IMF.

The simplicity of a universal IMF faces, however, serious challenges. From a theoretical perspective, it remains unclear why the IMF should or should not depend, e.g., on the local conditions of the gas out of which new stars are formed \citep{Hennebelle08,Myers11,Hopkins12IMF,Krumholz14,Chabrier14,Fontanot18,Guszejnov19,Davis20,Sharda22,Tanvir22}. Furthermore, even within the Milky Way, systematics deviations from a universal IMF have been claimed in order to explain the observed properties of galactic globular clusters \citep{Marks}. 

Beyond the limited range of star formation conditions probed by the Milky Way environment, ETGs are ideal systems to assess the universality of the IMF. Their old stellar populations are relatively simple \citep[e.g.][]{Faber73,Tinsley76,vazdekis:97,Trager00,Thomas05} and their high luminosities and well-behaved radial light distributions allow for precise analyses of their internal dynamics \citep[e.g.][]{Jeans22,Schwarzschild79,Remco08,Cappellari08,Ling18b}. Both stellar population \citep[e.g.][]{vandokkum,spiniello12,Spiniello2013,Spiniello15,conroy12,Smith12,ferreras,labarbera,Tang17,Lagattuta17,Rosani18,Lonoce21,Lonoce23} and dynamics-based \citep[e.g.][]{Treu,thomas11,auger,cappellari,Dutton12,wegner12,Tortora13b,Lasker13,Tortora14,Posacki15,Li17,Corsini17,Sonnenfeld19} IMF measurements in massive ETGs systematically point towards an excess of low-mass stars in these objects compared to the expectations from a Milky Way-like IMF. While there are systematical differences between both approaches \citep{Smith13,smith,Lyubenova16,Davis17,MN19,Lu23}, these results suggest that the extreme condition under which stars in massive ETGs formed altered the fragmentation of molecular clouds, leading to the observed IMF variations.

Yet, although central to our current understanding of the IMF, massive quiescent ETGs can only provide information about the IMF over a very limited stellar mass range, from the Hydrogen burning limit to $\sim$1\msun, since stars more massive than that have long evolved into dark remnants. Measurements of the high mass end of the IMF in young stellar populations have been generally limited to emission-line studies, in particular comparing the ionizing flux of massive stars to the overall stellar continuum \citep[e.g. UV flux or optical colors][]{Hoversten08,Meurer09,Lee09,Nanayakkara17}, providing tentative evidence of an IMF deficient in high-mass stars in dwarf galaxies. Complementary, submillimetric measurements \citep[e.g.][]{Romano17} of nearby and distant star forming galaxies suggest that the IMF in these relatively massive objects is biased towards massive stars \citep{Sliwa17,Brown19,Zhang18}, leaving the door open to an IMF simultaneously bottom and top-heavy in massive galaxies \citep[e.g.][]{Ferreras15}.

In this paper we present an alternative approach to constraining the IMF in young stellar populations based on the analysis of absorption spectra, building up on the success of stellar population synthesis models in revealing the variable nature of the IMF in massive quiescent galaxies. We apply this novel approach to a nearby late-type galaxy, NGC\,3351, drawn from the Time Inference with MUSE in Extragalactic Rings (TIMER) survey \citep{TIMER}, obtaining a spatially resolved two-dimensional map of the IMF in this galaxy. While IMF maps have been already derived for several nearby galaxies \citep[][]{MN19,MN21,Barbosa20}, this is the first time that this has been achieved for a young star-forming system. The layout of this paper is as follows: \S~\ref{sec:data} presents the TIMER data of NGC\,3351. In \S~\ref{sec:model} we give a detailed explanation of our stellar population modeling approach. The results of our analysis of NGC\,3351 are presented in \S~\ref{sec:results} and discussed in \S~\ref{sec:discussion}. Finally, \S~\ref{sec:summary} summarizes our findings and lists our concluding remarks. 

\section{Data} \label{sec:data}

The TIMER project presented in \citet{TIMER} is a MUSE survey of 24 nearby barred galaxies. The exquisite quality and depth of the TIMER data, exemplified in some of its most recent results \citep[e.g.][]{Adri19,jairo19,Gadotti20,Justus2020,Bittner21} are ideally suited to measure the subtle effect that a variable IMF has on the integrated spectra of galaxies. These data were obtained through the MUSE integral field unit \citep{Bacon10}, covering a wavelength range from $\sim$4700 \AA \ to 
$\sim$ 9000 \AA \ at a mean resolution of R$\sim$3000. A detailed description of the sample and data properties are given in the survey presentation paper \citep{TIMER}.

\begin{figure}
   \centering
   \includegraphics[width=9cm]{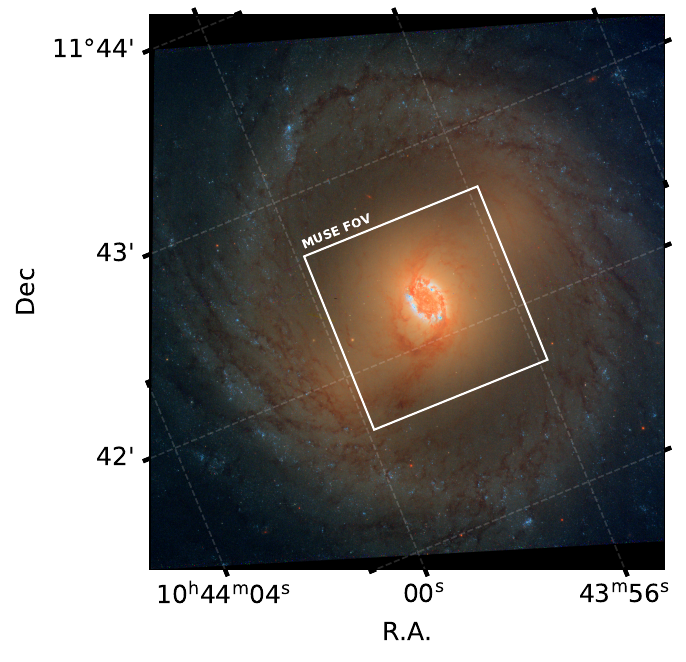}
   \caption{HST color image of NGC\,3351. The star forming ring is clearly visible in this color-composite image in the inner regions of NGC\,3351, while the presence of a long bar becomes also clear as an elongated structure that runs diagonally across the HST image, extending beyond the MUSE FOV indicated with a white square (1'x1').}
   \label{fig:rgb}
\end{figure}

This paper focuses on the late-type SB galaxy NGC\,3351, with an estimated stellar mass of \hbox{M$_\star=3.1 \times 10^{10}$ \msun} \citep{S4G} and a distance of 10.1 Mpc\footnote{At this distance, the pixel scale of the MUSE spectrograph translates into a spatial scale of $\sim 10$ pc/spaxel}. Fig.~\ref{fig:rgb} shows a HST color image of NGC\,3351 (F438W, F555W, F814W)\footnote{Program ID 15654, PI Janice Lee}, where its main morphological features are clearly visible, including the star-forming ring and the bar, extending diagonally across the HST image and beyond the MUSE field-of-view (FOV). In the context of the TIMER survey, \citet{Leaman19} proposed that the observed (ionized and molecular) outflow in NGC\,3351 is powered by the energy and momentum injected by its star forming ring. Moreover, \citet{Bittner20} reported that the stellar populations in this star forming ring have a relatively high [Mg/Fe] ratio and low metallicity but the origin of such a potentially chemically peculiar gas remains unknown. The relatively high mass and thus luminosity of NGC\,3351 and its peculiar stellar population content makes it an ideal test case for our pilot study.

Although as indicated above the spatial resolution of the MUSE data is of $\sim 10$ pc/spaxel, measuring the possible effect of the IMF requires high signal-to-noise ratio spectra. Therefore, instead of analyzing the stellar population properties on a spaxel-by-spaxel basis, we used the Voronoi binning code presented in \citet{voronoi} to spatially combine nearby spaxels in order to achieve a minimum signal-to-noise ratio per \AA \ of 100. This binning procedure effectively decreases the spatial scales we can resolve but ensures that the individual spectra to be fit are deep enough for our detailed stellar population analysis. All the results presented in this paper will be therefore derived from this Voronoi-binned MUSE data cube.  

\section{Stellar population modeling} \label{sec:model}

\subsection{Stellar population synthesis models}
Our approach is based on the MILES stellar population synthesis models \citep{miles}. In particular, we use the $\alpha$-variable models presented in \citet{Vazdekis15} in order to account for possible variations in the abundance pattern. These models are built by combining the BaSTI set of isochrones \citep{basti1,basti2} with the MILES stellar library \citep{Pat06} and the synthetic stellar library of \citet{Coelho05}. The main advantage of these models is their consistent and semi-empirical treatment of $\alpha$-elements variations, providing intermediate resolution model spectra from $\sim$3500\AA \ to $\sim$7400\AA \ \citep[see also][]{Walcher09}.

The $\alpha$-variable MILES models cover a range of total metallicity from [M/H]=-2.27 to [M/H]=+0.40, and from 30 Myr to 14 Gyr in age. All $\alpha$-elements are varied simultaneously and model predictions are given at [$\alpha$/Fe]=+0.0 and [$\alpha$/Fe]=+0.4. To model possible IMF variations, we assume in this work a single power law IMF parametrization (the so-called {\it unimodal} IMF shape in the MILES notation, \citealt{vazdekis96}). Changes in the IMF are parametrized by  $\Gamma$, the logarithmic slope of the IMF. For reference, a Salpeter IMF \citep{Salp:55} corresponds to $\Gamma=1.35$. Note that other IMF functional forms such as log-normal approximation \citep[e.g.][]{Chabrier} or a broken power law \citep[e.g.][]{kroupa} have no exact translation into $\Gamma$ units.

It is worth mentioning that the here-adopted single power law IMF description departs from our previous measurements of the IMF in ETGs galaxies. Observations in the Milky Way \citep[e.g.][]{MW,bastian} and theoretical arguments \citep[e.g.][]{Chabrier14,Krumholz11,Krumholz14} support the presence of a characteristic stellar mass in the IMF and therefore strongly disfavor an IMF defined by a single slope across all masses. Moreover, mass-to-light ratio predictions for a single power law IMF in massive ETGs are inconsistent with dynamical mass measurements \citep{ferreras,Cappellari13}. However, given our current lack of knowledge regarding the IMF behavior in young stellar populations beyond the Milky Way, a single power law IMF parametrization greatly simplifies the interpretation of the obtained results.

Young systems with extended star formation histories can, however, contain stellar populations younger than the limit of 30 Myr of the MILES $\alpha$-variable models. This can be particularly relevant for the spatially resolved IFU data of the TIMER survey since, even if the average age of a galaxy is moderately old, star formation can be restricted to small regions where populations younger than 30 Myr can in fact significantly contribute to the observed spectra. To overcome this potential issue, our analysis complements the $\alpha$-variable models with the super-young MILES models presented in \citet{youngMILES}. Fed with the \citet{Padova94} set of isochrones, these super-young MILES models provide predictions down to 6.3 Myr.

\begin{figure*}
   \centering
   \includegraphics[width=18cm]{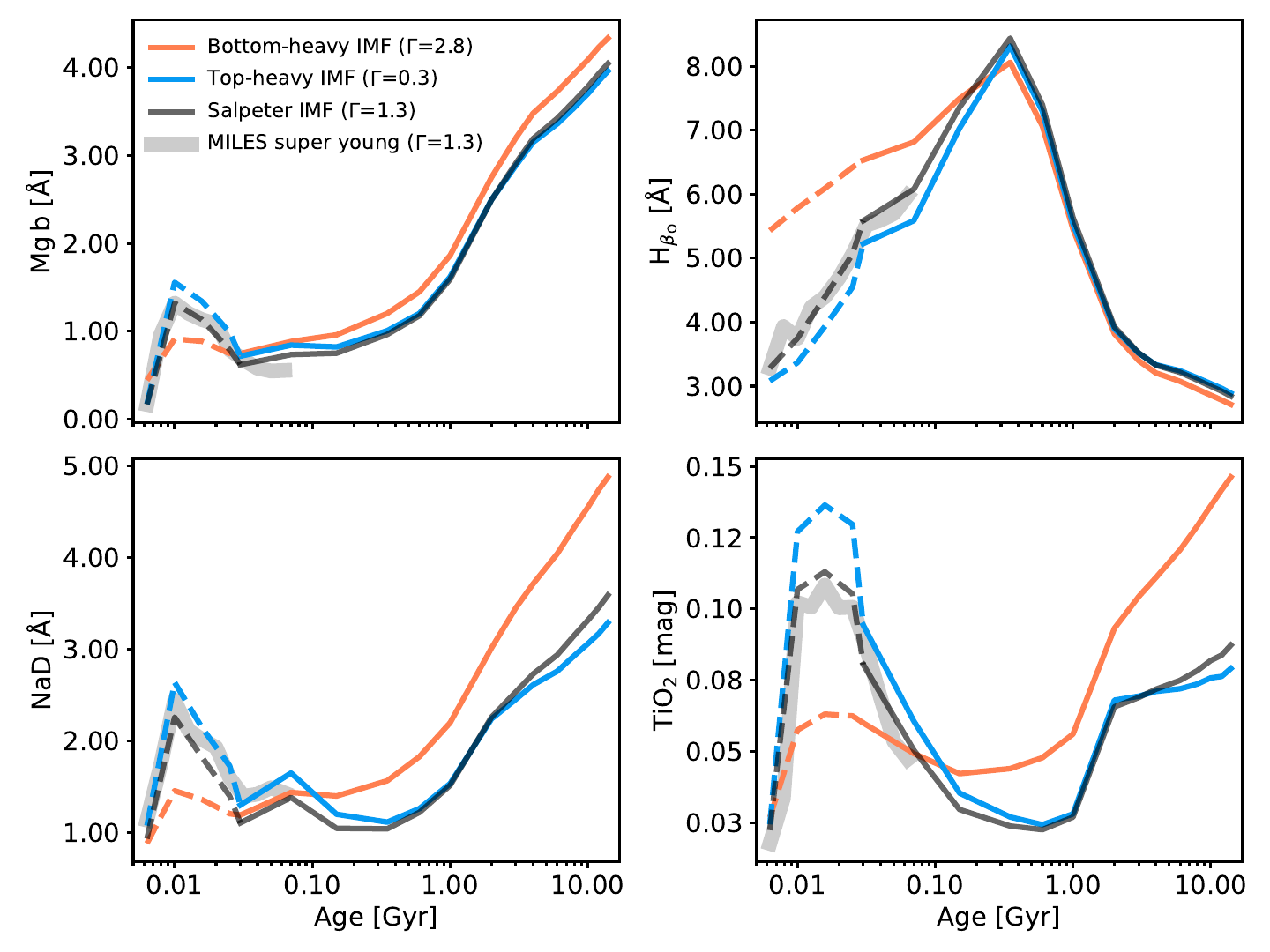}
   \caption{Line strength age and IMF sensitivity. From left to right and top to bottom we show the age dependence of the Mgb, H$_{\beta_O}$, Nad, and TiO$_2$ spectral features, respectively. Solid lines correspond to the MILES $\alpha$-variable models and dashed ones indicate their (adapted) super-young extension. The gray shaded area in the background shows the unmodified super-young model predictions for a Milky Way-like IMF (see main text for more details). Orange lines are predictions for a bottom-heavy IMF (i.e. steeper than the Milky Way standard), blue lines for a top-heavy (i.e. flatter) IMF, and black lines for the reference Salpeter IMF. The rapid increase in the equivalent width of the indices (but H$_{\beta_O}$) at ages around $\sim 10$ Myr is caused by the sudden appearance of red supergiant stars. Predictions are shown at the native resolution of the MILES models (2.51 \AA).}
   \label{fig:indices}
\end{figure*}

In order to combine both $\alpha$-variable and super-young MILES models into a regular grid of spectra, we linearly interpolated the latter to match the metallicity sampling of the $\alpha$-variable models. Moreover, although the effect of a variable abundance pattern becomes weaker as the age of the stellar population decreases \citep[see e.g. Figures 9 and 10 in][]{Vazdekis15}, we used the behavior of the $\alpha$-variable MILES models to expand our treatment of the abundance pattern towards the super-young models. In practice this was done by, at a given metallicity, calculate the ratio of the [$\alpha$/Fe]=+0.0 and [$\alpha$/Fe]=+0.4 models for the youngest $\alpha$-variable models. This response function was then applied to the super-young models in order to approximate the effect of a variable abundance ratio. Note that, since these super-young models are based on the MILES stars that follow the local [$\alpha$/Fe]-[Fe/H] relation \citep[see][]{Milone11,MILES21}, the correction of the super-young models depends on the metallicity. Below [M/H] = -0.4, models have effectively [$\alpha$/Fe]$\sim0.4$ and therefore we use the response functions to calculate model predictions at [$\alpha$/Fe]$\sim0.0$. Conversely, above [M/H] = -0.4 we correct the super-young models to reach [$\alpha$/Fe]$\sim0.4$.

In short, after combining the $\alpha$-variable and super-young MILES models, our set of simple stellar population (SSP) model predictions cover ages from 6.3 Myr to 14 Gyr, with variable total metallicity, [$\alpha$/Fe], and IMF, this one parametrized as a single power law with logarithmic slope $\Gamma$.  

\subsection{Line-strength behavior}

Measuring the IMF from integrated spectra requires probing the contribution to the observed flux of stars with different masses. From a modelling point of view, this is done by mapping (at fixed chemical composition) effective temperature and surface gravity onto stellar mass across a given isochrone. IMF-sensitive features are therefore usually labelled as either \textit{temperature-sensitive} or \textit{gravity-sensitive}. Note, however, that the net effect of a variable IMF on a given spectral feature is the result of a (non-trivial) combination of its sensitivity to both temperature and surface gravity, modulated by the shape of the isochrone, the intrinsic mass-luminosity relation of the stars, and the adopted IMF. The titanium dioxide molecular band TiO$_2$ at $\sim6200$ \AA \ \citep{Worthey94b} is a paradigmatic example of this complex IMF sensitivity, as for old stellar populations it becomes stronger when increasing the number of dwarf stars (i.e. with bottom-heavier IMFs) even though it is actually more prominent in the atmospheres of giants \citep[see e.g. Figure 1 in][]{Spiniello2013}.

Therefore, before further discussing our fitting approach, it is worth familiarizing with the behavior of some of the most prominent spectral features within the MUSE wavelength range, particularly for young ages, where their IMF sensitivity remains virtually unexplored. Figure~\ref{fig:indices} shows how the Mgb, NaD, TiO$_2$ \citep{Worthey94b} and H$_{\beta_O}$ \citep{Cervantes} line-strength indices change as a function of age, for three different IMF slopes (at solar metallicity). Ages where the model predictions come from the super-young MILES models are indicated by a dashed line. 

Arguably, the most characteristic feature in Fig.~\ref{fig:indices} is the predicted bump in all indices but H$_{\beta_O}$ for ages around $\sim 10$ Myr. Although it roughly coincides with the transition towards the super-young models, the rapid change in the indices strength is not an artifact of joining the two sets of models but a consequence of the onset of the red supergiant phase \citep[e.g.][]{Mayya97}. This is in fact shown by the fact that the gray shaded areas in Fig.~\ref{fig:indices}, indicating the raw super-young MILES models predictions (i.e., without the metallicity and [$\alpha$/Fe] corrections detailed above), closely track the behavior of the black lines, both calculated for the same Salpeter IMF. 

The importance of including these super-young model predictions in our analysis becomes obvious from Fig.~\ref{fig:indices}, particularly for the case of the TiO$_2$ feature. In the absence of these models, the expected contribution of red supergiants to the observed spectra would be severely underestimated and one could wrongly interpret a strong TiO$_2$ absorption as a consequence of a dwarf-rich (i.e. bottom-heavy IMF) while in reality results from the presence of very young stellar populations. It is also important to notice that ages in Fig.~\ref{fig:indices} are shown in logarithmic scale and therefore the actual dominance of red supergiants is effectively restricted to a very short phase peaking at $\sim 10$ Myr. Our fitting procedure will take advantage of this rapid evolution (see \S~\ref{sec:fit})

The model predictions represented in Fig.~\ref{fig:indices} also highlight important differences in the IMF sensitivity of these indices depending on the age of the underlying stellar populations. Both TiO$_2$ and NaD absorption features have been extensively used to characterize the IMF in old, quiescent galaxies with a rather straightforward interpretation: stronger features translate into steeper IMF slopes. This, however, does not hold for younger stellar populations where the opposite trend is expected. This reversal of the IMF dependence is clearly exemplified by the TiO$_2$ feature where similarly strong values are predicted for either an old population and a dwarf-dominated IMF or a young population biased towards massive stars. 

A final note regarding the effect of a variable IMF on the spectra of young stellar populations. Having young populations implies that more massive stars still contribute to the observed spectra. This has two immediate and profound consequences for the interpretation of the measurements. First, contrary to the old stellar populations found in quiescent galaxies where only stars less massive than $\sim1$ \msun \ are present, IMF measurements in young populations are {\it age-dependent} since the stellar mass range probed along the IMF changes as populations become younger. Thus, changes in the measured slope might be observed, e.g. within the same galaxy, not because the IMF itself is varying, but because stellar populations with different ages probe different mass ranges across the IMF, each of them characterized by a different slope.

\begin{figure}
   \centering
   \includegraphics[width=9cm]{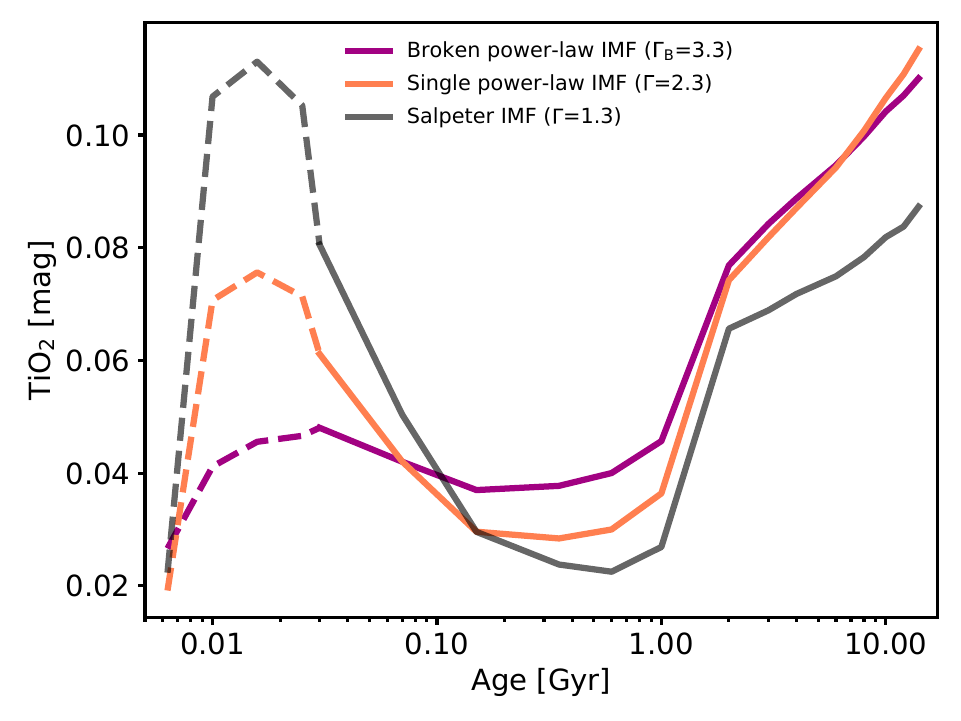}
   \caption{Optical sensitivity to the shape of the IMF. Black, orange, and purple lines indicate the age sensitivity of the TiO$_2$ absorption feature for a Salpeter IMF, a bottom-heavy single power law IMF, and a bottom-heavy broken power law IMF, respectively. For old stellar populations, as those harbored by massive quiescent galaxies, both single power law and broken power law IMF parametrizations are indistinguishable since the stellar mass range probed along the IMF is rather narrow. For young stellar populations however, different model assumptions on the IMF shape lead to dramatically different model predictions, highlighting the potential use of line-strength indices to actually constrain the shape of the IMF. Predictions are shown at the native resolution of the MILES models (2.51 \AA).}
   \label{fig:shape}
\end{figure}

Second and related to the point above, in old stellar populations, IMF measurements are limited to a narrow mass range, resulting in a lack of sensitivity to the  IMF parametrization. This has greatly simplified the comparison between different observational approaches and IMF variations are reliably discussed in terms of a single parameter (e.g. IMF slope $\Gamma$, dwarf-to-giant ratio F$_\mathrm{0.5}$, mismatch parameter $\alpha$, etc.). The spectrum of a young stellar population,  however, becomes progressively sensitive to the actual shape of the IMF, as demonstrated by Fig.~\ref{fig:shape}, where te equivalent width of the TiO$_2$ IMF sensitive feature is shown as a function of age, for different IMF functional forms.

Fig.~\ref{fig:shape} shows predicted equivalent width of the TiO$_2$ as a function of age for three different IMFs and solar metallicity. As in Fig.~\ref{fig:indices}, orange and black lines correspond to a bottom-heavy single power law and a Salpeter IMF, respectively. In purple we show the predictions for a bottom-heavy IMF (i.e. biased towards low-mass stars), but assuming a broken power law mass distribution \citep[the so-called {\it bimodal} IMF within the MILES notation, ][]{vazdekis96}. This {\it bimodal} IMF aims to be a generalization of the \citet{kroupa} functional form, with a variable slope for the high-mass end $\Gamma_\mathrm{B}$ and a fixed, flat distribution for low mass stars (below $\sim0.4$\msun). For $\Gamma_\mathrm{B}$, the {\it bimodal} IMF is equivalent to the Milky Way standard.

For old ages, the predicted TiO$_2$ for the two bottom-heavy IMF parametrizations are virtually indistinguishable. This is ultimately due to the fact that in old populations, spectra only probe a very narrow stellar mass range, resulting in a lack of sensitivity to the IMF functional form \citep{labarbera}. However, as populations get younger, stars with different masses start to contribute to the observed spectra and line-strength analyses become sensitive to the shape of the IMF. Finally, and in the absence of more flexible and physically-motivated functional forms, the trends in Fig.~\ref{fig:shape} also justify our adoption of a single power law IMF parametrization. Although we acknowledge that such a functional form is inconsistent with dynamical IMF measurements of massive quiescent galaxies \citep[e.g.][]{ferreras,Lyubenova16}, it is relatively agnostic about the physics behind IMF variations and it has a simple interpretation even for populations with different ages.

\subsection{Limitations of the SSP assumption}

The effect of the IMF on the integrated spectra of galaxies is subtle and it can be partially mimicked by changes in other stellar population parameters such as the chemical composition or the star formation history. To overcome this problem, IMF measurements based on the analysis of absorption spectra have been so far limited to quiescent galaxies whose star formation histories are reasonably well-represented by SSP models. The SSP assumption simplifies the fitting process by reducing the number of free parameters and isolating the effect of the IMF becomes feasible.

\begin{figure*}
   \centering
   \includegraphics[width=9cm]{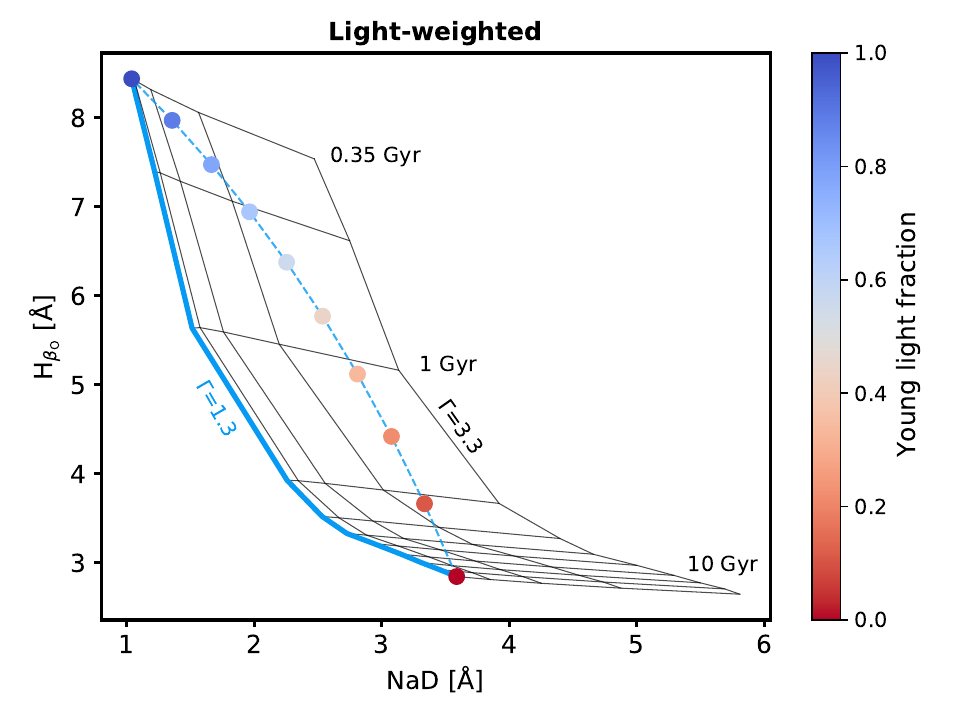}
   \includegraphics[width=9cm]{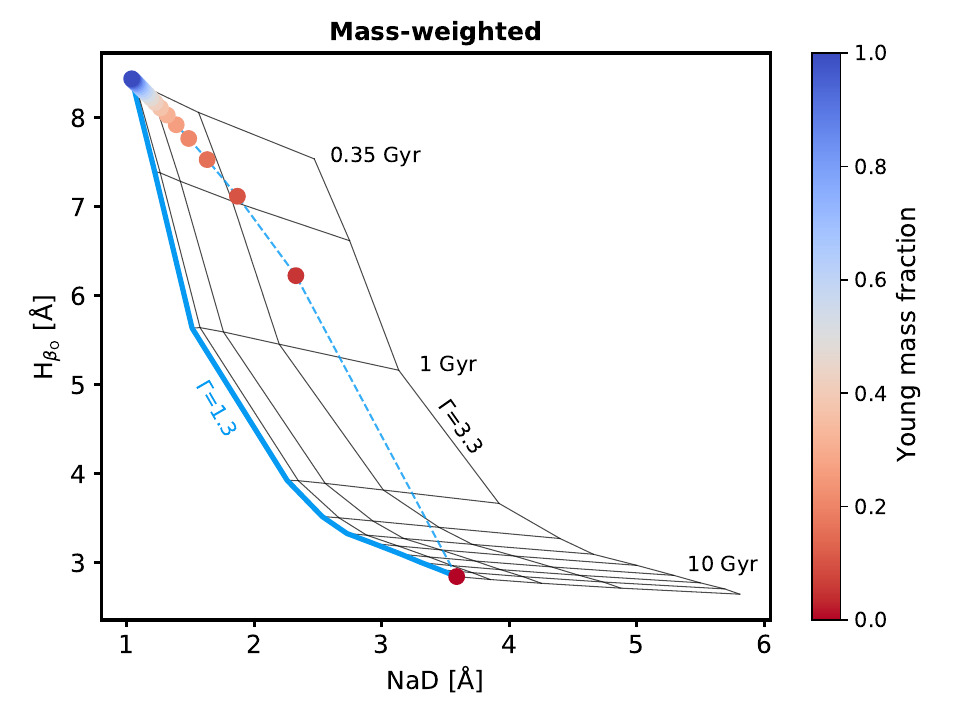}
   \caption{H$_{\beta_o}$ vs NaD index-index diagram. Gray grids show the MILES predictions at the native resolution of the models (2.51 \AA) and solar metallicity for different ages and IMF slopes, as indicated by the labels. Blue solid lines highlight the age variation for a Salpeter IMF. Dashed blue lines and filled circles track the range of expected values for a composite stellar population of two SSP models (14 Gyr and 0.35 Gyr), both of them with a Milky Way-like IMF. Symbols are color-coded by the light (left panel) and mass (right panel) fraction associated to the young SSP model. This simplified yet realistic example shows how the best-fitting SSP solution fails to recover the true value of $\Gamma=1.3$ when applied to composite stellar populations, which could result in unrealistically high IMF slope values ($\Gamma\sim 3.3$).}
   \label{fig:offgrid}
\end{figure*}

Although it has been central to our current understanding of the stellar populations in galaxies, the SSP assumption faces unavoidable problems when dealing with systems with more complex star formation histories \citep[e.g.][]{Gallazzi05,Serra07,Walcher15}. The limitations of the SSP approach are illustrated in Fig.~\ref{fig:offgrid}. The two panels in this figure show a standard index-index diagram with an optimized age-sensitive feature in the vertical axis \citep[H$_{\beta_o}$,][]{Cervantes} and an IMF-sensitive feature on the horizontal one \citep[NaD,][]{Worthey94b}. Each corner of the grid corresponds to a MILES model prediction with a given IMF slope $\Gamma$ and age, as indicated by the labels. Over-plotted on top the index-grid, colored symbols indicate the expected H$_{\beta_o}$ and NaD values after combining two SSP models of 0.35 and 14 Gyr with exactly the same $\Gamma=1.3$ Milky Way-like IMF slope. Each of these symbols is color-coded by the fraction of light (left) or mass (right) associated to the young SSP model. 

Two main conclusions can be directly drawn from Fig.~\ref{fig:offgrid}. First, the IMF of composite stellar populations should not be studied under the SSP assumption. In the simplified case illustrated in Fig.~\ref{fig:offgrid}, when the relative flux emitted by the old and young populations is similar, the best-fitting SSP solution for the IMF slope becomes close to $\Gamma=3.3$, failing to recover the true value of $\Gamma=1.3$. Note that the unreliability of the SSP assumption for composite stellar populations also does not only apply to IMF variations, but it also invalidates measurements of more robust quantities such as metallicities and elemental abundance ratios \citep[see e.g. Appendix~B in ][]{Seidel15}. Second, the SSP approach works best for young stellar populations. This, at first, may seem counterintuitive since the concept of SSP has been traditionally used in the context of old, quiescent stellar populations. However, young stellar populations can easily outshine the presence of any old component, as shown on the right panel of Fig.~\ref{fig:offgrid}. Modelling the flux emitted by young populations is obviously not free from uncertainties and systematics \citep[see e.g.][]{Leitherer99,Leitherer14,Eldridge17} but, given a set of SSP templates, the importance of accounting for an extended star formation history is maximal when old and young stellar populations contribute similarly to the observed flux budget. 

These limitations of the SSP approach can be directly tackled by allowing for an extended star formation history when fitting the observed absorption spectra of galaxies. This is the foundation of our stellar population fitting scheme described in detail in the section below. 

\subsection{Fitting scheme} \label{sec:fit}

An ideal stellar population fitting algorithm would be able to retrieve, from an integrated spectrum, the time evolution of every stellar population property. This would include, not only the star formation history, but also the time evolution of the chemical composition and, potentially, that of the IMF. However, and despite significant advances in measuring time-varying properties of galaxies from their integrated spectra \citep[see e.g.][]{CF05,Ocvirk06,Vespa,Cappellari17,Wilkinson17,prospector}, it remains unclear to what extent all that information can be recovered or if it is actually even sufficiently encoded in absorption spectra \citep{Ferreras22}.

As a first step forward, in this work we build upon our approach to measure the IMF described in \citet{MN19,MN21}. Starting with the Voronoi-binned MUSE data described above, our stellar population analysis consists of two steps. During the first one, we use the full spectral fitting code pPXF \citep{ppxf} to measure the kinematics (V and $\sigma$) of every spectra and to model and subtract the ionized gas emission from the observed spectra. Then, fixing the measured kinematics to avoid degeneracies with the stellar population parameters \citep{Pat11}, we re-run pPXF but this time imposing a regularized solution to the best-fitting weights distribution \citep{Cappellari17}. As detailed in \citet{MN19}, the main goal of this second pPXF run is to estimate the star formation history and thus, the mean age of each Voronoi-binned spectra. Calculating the age in this way has three main advantages. First, we minimize the age - abundance pattern degeneracy that arises from the H$_{\beta}$ dependence on [C/Fe] \citep[e.g.][]{conroy,LB16} and the lack of strong C-sensitive features within the MUSE wavelength range. Second, we robustly measure the age by simultaneously fitting a wide spectral range, from 4700 \AA \ to 6400 \AA. Finally and crucial to our approach, contrary to the standard line-strength analysis where only the zeroth order moment of the age distribution is measured, by using pPXF we can approximate the full star formation history of a given population. 

For consistency with the second step explained below, we use pPXF to measure luminosity-weighted quantities. In practice, and related to what it is shown in Fig.~\ref{fig:offgrid}, we calculate the light fraction contained in stars with different ages. We regularize the pPXF solution in the age-metallicity-IMF space which, minimizing the dependence of the recovered star formation history on the assumed IMF. In order to estimate the uncertainty in the measured star formation history, we repeat the analysis of each spectrum ten times, adding Gaussian noise to the observed spectra with an amplitude given by the residuals of the best-fitting model. The end result of this first step is therefore that, for each spectra, we measure its kinematics, we remove the gas emission, and we estimate the star formation history and its uncertainty. This first part of the fitting scheme is the same as in \citet{MN19,MN21} and \citet{Bittner20} and it has proven to be robust and efficient to provide detailed stellar population maps based on MUSE data.

\begin{figure*}
   \centering
   \includegraphics[width=14cm]{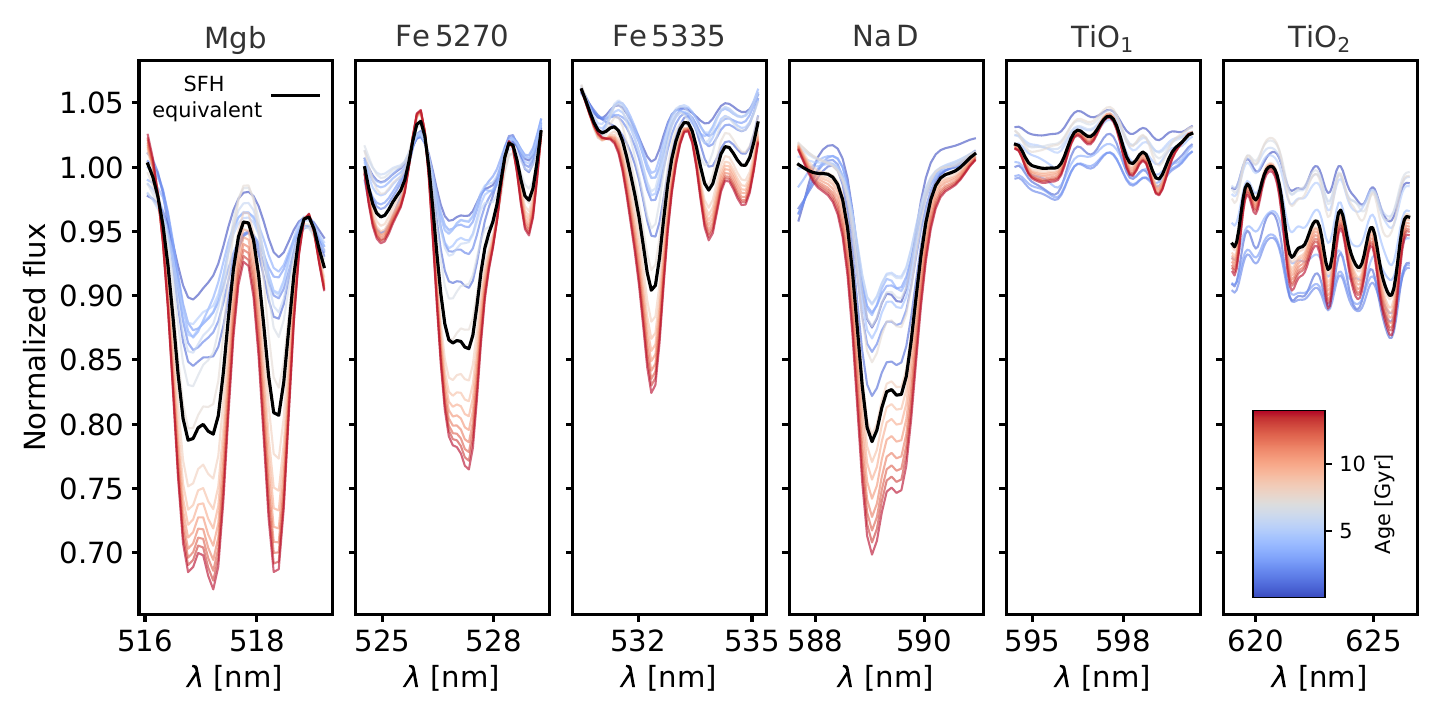}
   \caption{Colored lines illustrate the age dependence of the six spectral features included in our analysis. The solid black line shows the {\it SFH-equivalent} prediction resulting from weighting each SSP model according to the measured SFH. Note that the continuum of each model has been normalized to better represent the flux of the different SSPs. During the fitting process, however, the normalization is only applied to the SFH-equivalent model prediction (i.e. the black line).}
   \label{fig:combi}
\end{figure*}

During the second step, the goal is to measure the rest of the relevant stellar population parameters, including the slope of the IMF. Figure.~\ref{fig:offgrid} illustrates the limitations of the standard SSP approach that has been assumed so far in order to measure the IMF from integrated spectra. To overcome this problem, we have implemented a more flexible fitting scheme that allows for an extended star formation history while assuming a single chemical composition and IMF slope. In practice, the process is as follows. First, we use the star formation history derived from pPXF to generate a grid of MILES predictions. Note that this is a fundamental change compared to previous studies since the set of predictions are not SSP-like, but based on the measured star formation histories. In this way, by construction, we remove the bias shown in Fig.~\ref{fig:offgrid}. This combination of an extended star formation history and a single value characterizing the stellar population properties is in essence the same as most photometric inversion algorithms implement \citep[e.g.][]{Cunha,Kriek09,Han,prospector} and conceptually similar to the scheme followed by \citet{Gallazzi05}, but in our case optimized to measure the IMF from integrated absorption spectra. 

Given these star formation history-based MILES models predictions, we then use a Bayesian \citep{emcee} Full Index Fitting (FIF) approach \citep{MN19} to measure mean luminosity-weighted stellar population properties. In short, this is done by fitting every pixel within the band-pass of a set of spectral absorption features after normalizing their continuum. Summarizing the approach followed in this second step, Fig.~\ref{fig:combi} shows how SSPs with different ages have different equivalent widths and then contribute differently to the total flux of each spectral feature included in our analysis (see details below). The {\it SFH-equivalent} prediction is shown as a black line.

We select six prominent spectral features in the MUSE wavelength range, namely Mgb, Fe\,5270, Fe\,5335, NaD, TiO$_1$, and TiO$_2$. Modeling the band-passes of each of these indices based on the star formation history measured during the first step, our Bayesian FIF scheme fits for the following (luminosity-weighted) free parameters: total metallicity [M/H], [Mg/Fe], IMF slope, [Ti/Fe], and [Na/Fe]. Note that magnesium is effectively the only $\alpha$ element constrained by our set of indices through the Mgb feature. Therefore, even if MILES models are $\alpha$-variable, we refer to [Mg/Fe] when describing our results. Moreover, neither [Ti/Fe] nor [Na/Fe] abundances are self-consistently included in the MILES models. In particular, Ti is an $\alpha$ element but does not necessarily tracks Mg variations \citep[see e.g.][]{johansson12,Conroy14,Tang17}. In practice, the inclussion of a variable [Ti/Fe] allows us to approximate the effect of a variable abundance pattern in the behaviour of TiO mollecular bands as a nuissance parameter rather than constraining the actual [Mg/Fe] value. To account for the impact of [Ti/Fe] and [Na/Fe] we use the response functions of \citet{conroy}. These response functions are calculated for a 13 Gyr stellar population and since the sensitivity to elemental abundance variations becomes weaker for young populations, they resulting [Na/Fe] and [Ti/Fe] values are not reliable in absolute terms and will not be further discussed in this work. Motivated by the behavior of the indices in Fig.~\ref{fig:indices}, we include in our fitting scheme an additional free parameter that freely controls the fraction of flux emitted by a 10 Myr old population. In this way we can better model the critical contribution of supergiant stars. As noted above, the fact that this phase rapidly peaks at 10 Myr allows us to model it as a pure additional SSP contribution. Finally, we also include a re-scaling term to the estimated error in the MUSE data. Assuming uniform prior distributions for all free parameters, we repeat the fitting process ten times, one for each of the measured star formation histories (see above). We then combine all the posterior distributions to estimate the best-fitting values and their uncertainties. With all this, our stellar population analysis fits in total for seven free parameters: [M/H], [Mg/Fe], $\Gamma$, [Na/Fe], [Ti/Fe], F$_\mathrm{RSG}$, and $\Delta_\mathrm{err}$. 

\begin{figure*}
   \centering
   \includegraphics[width=14cm]{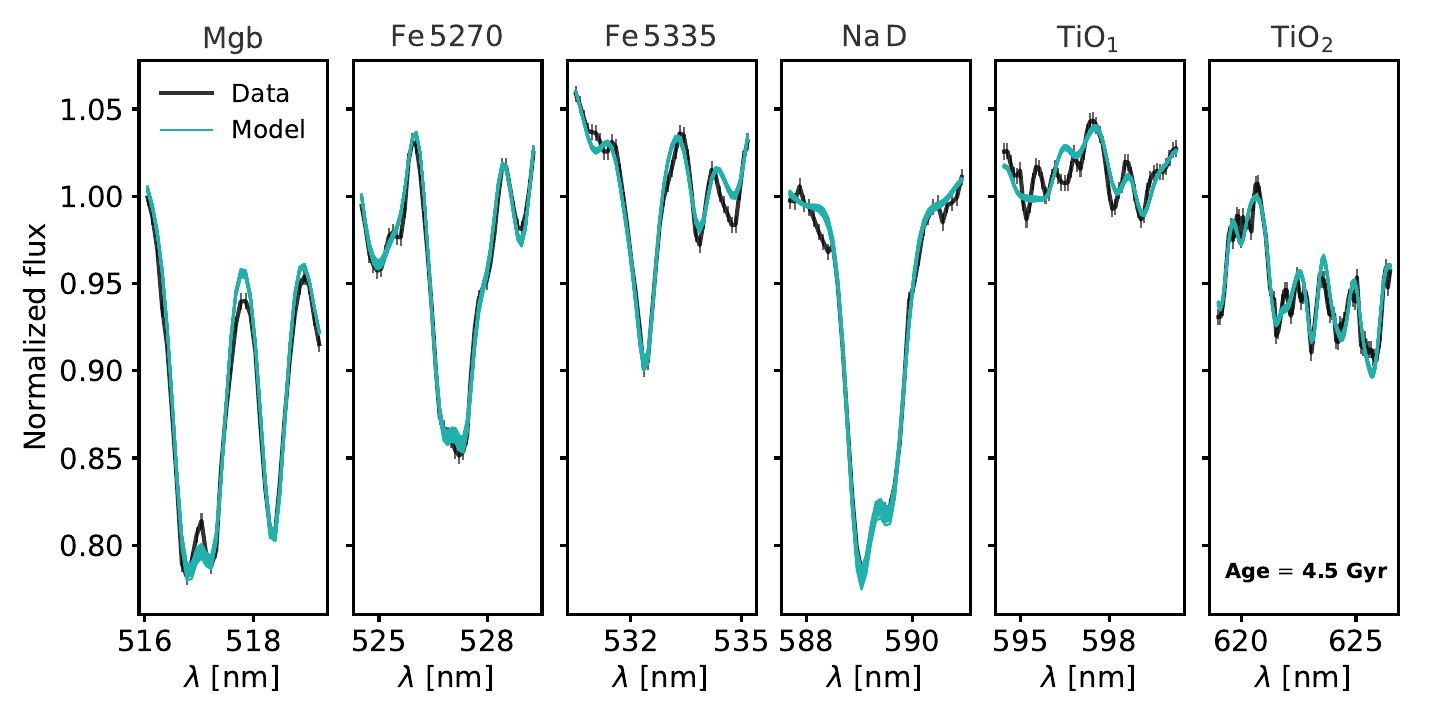}
   \caption{Data (in black) and best-fitting models (green) for one of the Voronoi-binned data of NGC\,3351. Each panel shows the central band-pass of the five spectral features used in our analysis: Mgb, Fe\,5270, Fe\,5335, NaD, TiO$_1$, and TiO$_2$. Each green line shows a best-fitting model resulting from sampling the full posterior distribution, taking into account the uncertainty in the recovered star formation history. The mean luminosity-weighted age is indicated on the rightmost panel (TiO$_2$) although the prediction for the index strength is calculated based on the measured star formation history (see text for more details).}
   \label{fig:fit}
\end{figure*}

\begin{figure}
   \centering
   \includegraphics[width=8.5cm]{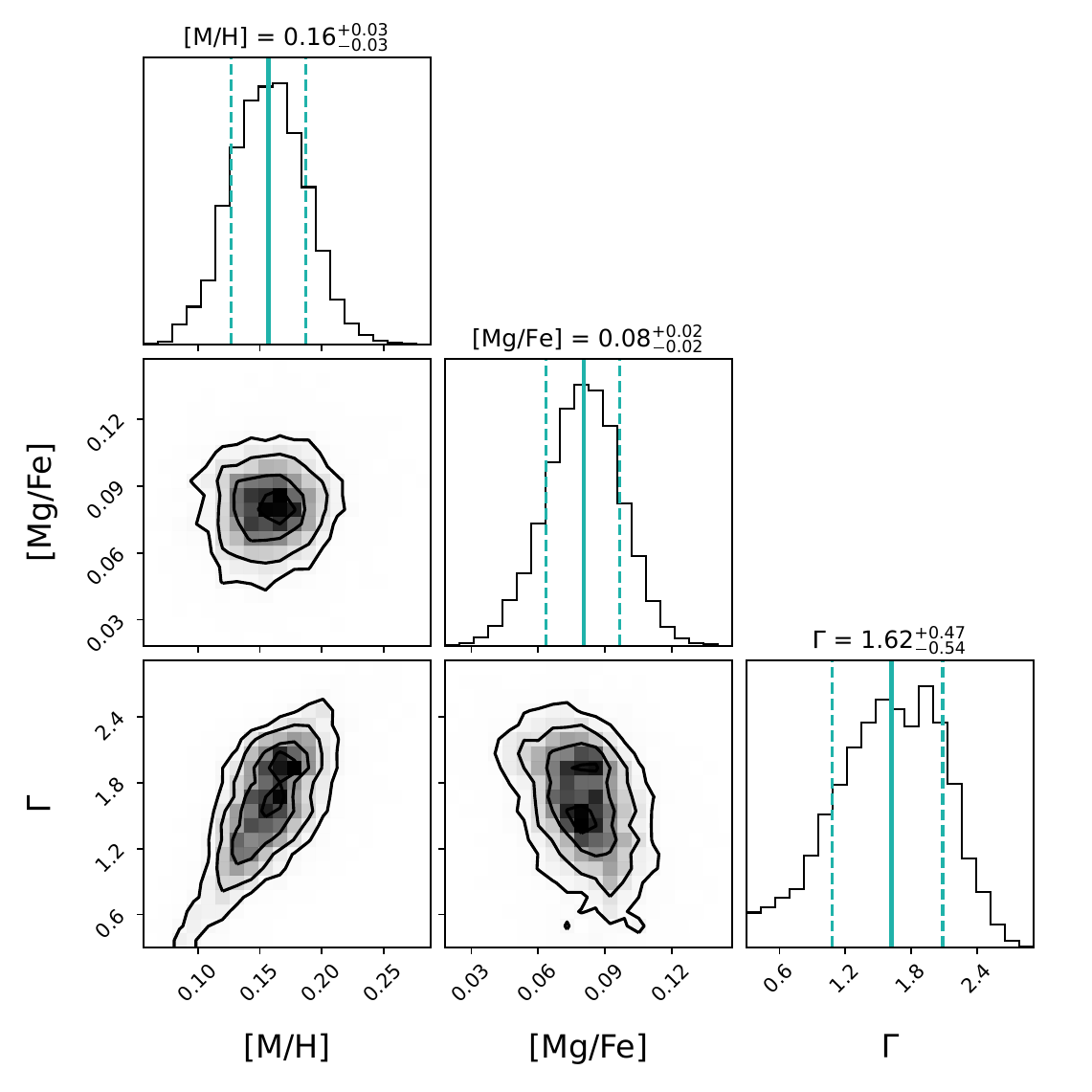}
   \caption{Full posterior distribution for the three main stellar population parameters measured during the second step, namely total metallicity, [Mg/Fe], and IMF slope $\Gamma$, for the observed spectrum shown in Fig.~\ref{fig:fit}. The green solid lines indicate the median of the distribution while dashed vertical lines mark the 16th and 84th percentiles. Our fitting approach is able to capture and successfully break the degeneracies of the inversion problem even when fitting stellar populations characterized by an extended star formation history.}
   \label{fig:corner}
\end{figure}

Figure~\ref{fig:fit} exemplifies the result of our fitting approach for one of the Voronoi-binned MUSE spectra of NGC\,3351. Note that the spectrum shown in Fig.~\ref{fig:fit} has a relatively young (luminosity-weighted) age as a result of an extended star formation history and our fitting scheme is able to reproduce the observed strength of all five indices in a regime where the SSP approach does not longer holds. The reliability of our approach is summarized in Fig.~\ref{fig:corner}, where the full posterior distribution of metallicity, [Mg/Fe] and IMF slope values is shown for the same spectrum as in Fig.~\ref{fig:fit}. Even in this non-trivial case with a young and extended star formation history, our fitting scheme is able to assess the intrinsic degeneracies of the inversion problem. A discussion on potential biases and systematics on the recovered stellar population properties, in particular on the IMF values, is included in Appendix~\ref{sec:system}.

\section{Results} \label{sec:results}

\begin{figure*}
   \centering
   \includegraphics[width=8cm]{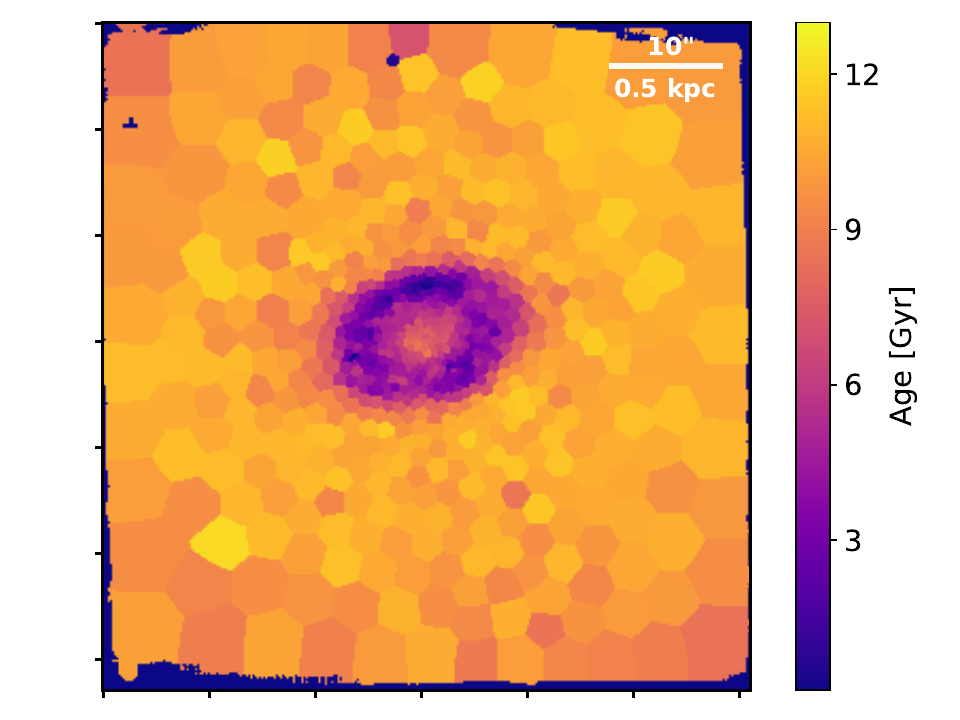}
   \includegraphics[width=8cm]{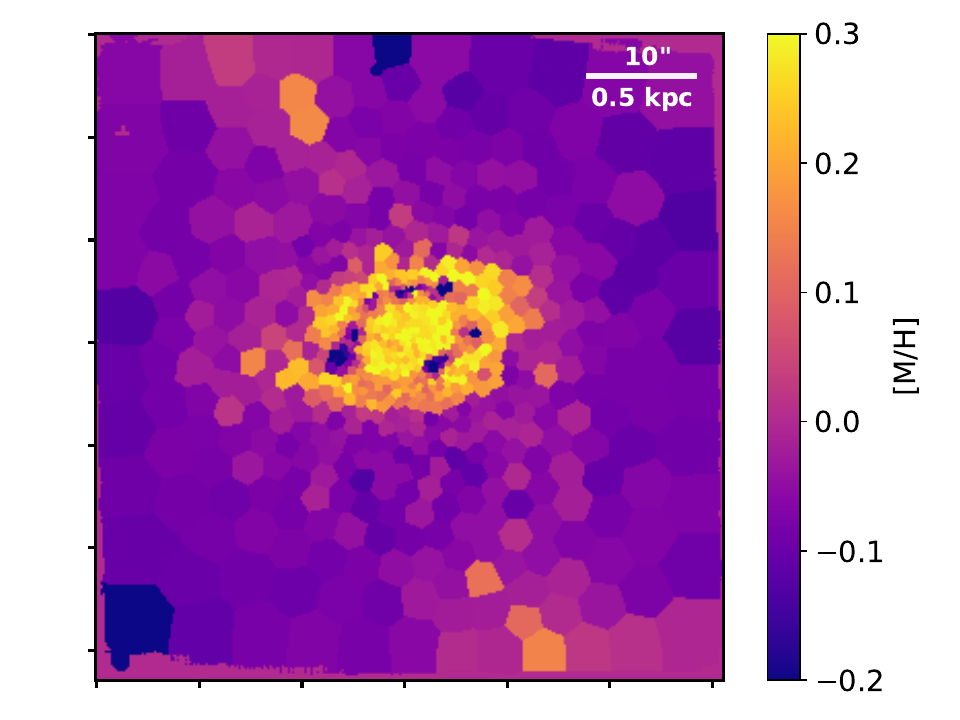}
   \includegraphics[width=8cm]{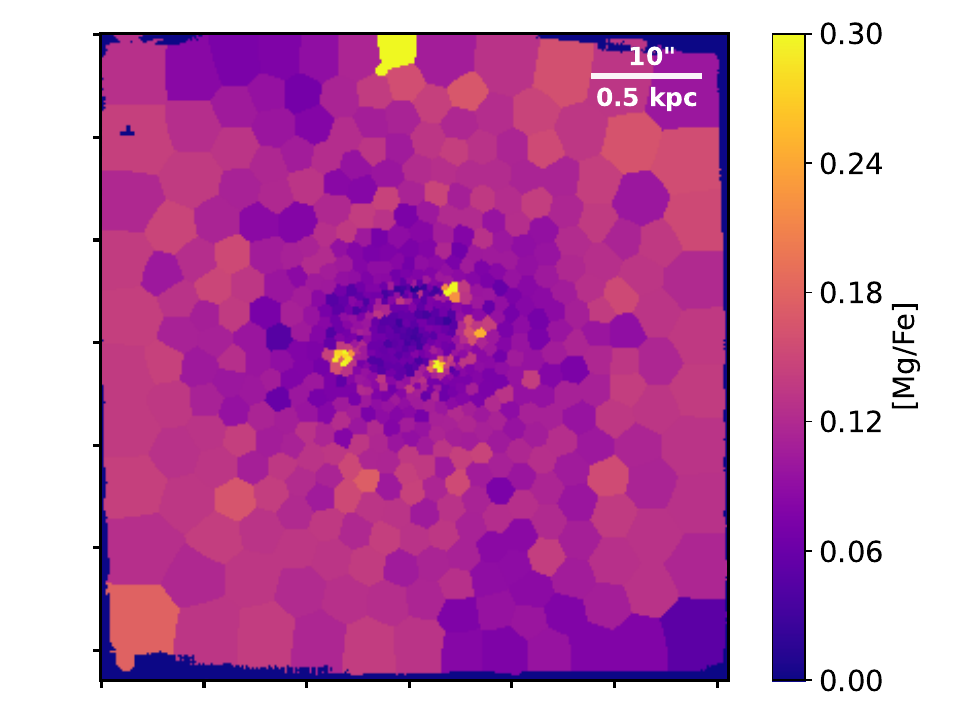}
   \includegraphics[width=8cm]{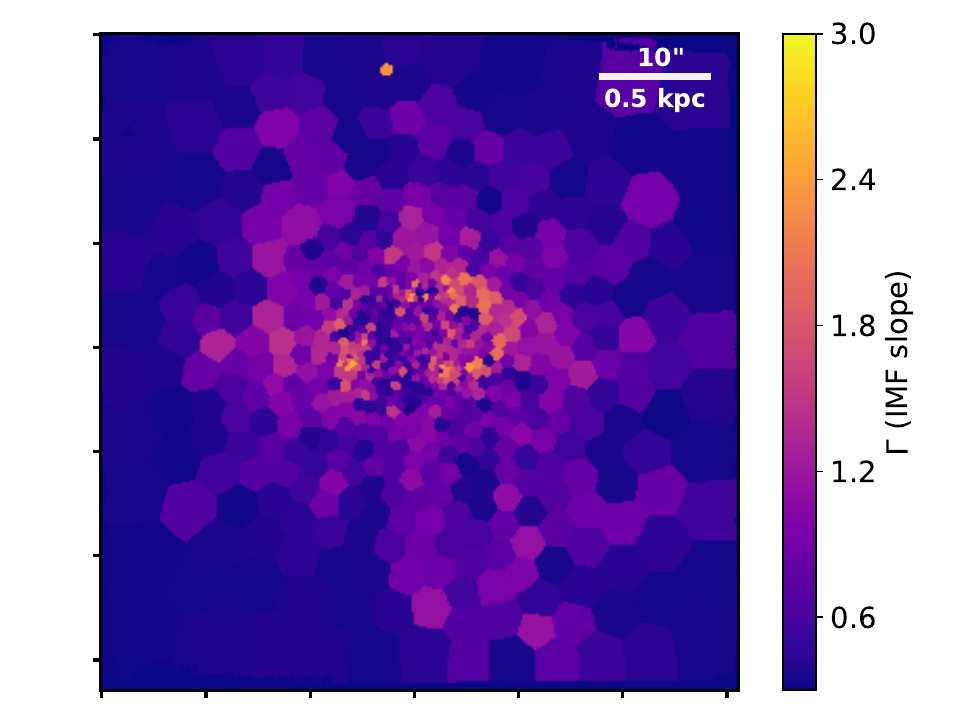}

   \caption{Best-fitting stellar population maps for NGC\,3351. The top left panel shows the age map, with a clearly visible nuclear star-forming ring characterized by the presence of young stellar populations. On the top right, the total metallicity map of NGC\,3351 shows the metal enrichment of both the nuclear disk and the bar. Regions of apparently low metallicity are visible across the star-forming ring. The [Mg/Fe] map (bottom left) exhibits a behavior similar (but inverted) to the metallicity map, with the nuclear disk and the bar appearing as low [Mg/Fe] components. Finally, the bottom right panel shows the IMF slope map where the typical values are close to what is observed in the Milky Way ($\Gamma = 1.3$). IMF variations, although mild, partially track changes in both metallicity and stellar density. For reference, a horizontal white line of 10" or 0.5 kpc at the distance of NGC\,3351 is shown in each panel.}
   \label{fig:maps}
\end{figure*}

Figure~\ref{fig:maps} shows the measured age, metallicity, [Mg/Fe], and IMF slope maps for NGC\,3351. The horizontal white lines indicate a scale of 10" or equivalently 0.5 kpc at the distance of the galaxy. The age map corresponds to the mean luminosity-weighted values measured during the first steep of the fitting process as described above and clearly reveals the star-forming nuclear ring of NGC\,3351 \citep[see][]{Leaman19, Bittner20}. Towards the outskirts, the stellar populations of NGC\,3351 become predominantly old. The prominent dust lines shown in Fig.~\ref{fig:rgb} do not have a clear impact on the age map, in agreement with the lack of (dust-corrected) H$_\alpha$ emission \citep{Bittner20}. Such dust lanes are known to lack star formation possibly due to the effects of shear on the molecular clouds \citep[e.g.][]{Lia92,Kim12,Emsellem15,Neumann19}.

The metallicity map is characterized by the presence of the metal-enriched nuclear disk. Almost perpendicularly, the bar of NGC\,3351 is clearly visible as an elongated structure that extends to the edge of the MUSE field of view. In addition, the metallicity map reveals the presence of what appear to be scattered metal-poor regions along the star-forming nuclear ring \citep{Bittner20,Pessa23}. These are star-forming regions with varying masses, dust attenuation levels, and formation histories as suggested by their photometric properties \citep{Calzetti21}. Interestingly, four [Mg/Fe] enhanced ([Mg/Fe]$ > 0.22$) regions are clearly visible in the map of NGC\,3351 shown in Fig.~\ref{fig:maps}. These chemically-peculiar (CP) regions, seemingly characterized by metal-poor and [Mg/Fe]-enhanced stellar populations, correspond to the youngest star-forming knots identified by \citet{Calzetti21}, with ages between $\sim5$ and $\sim10$ Myr. Beyond these CP regions, the [Mg/Fe] map shows rather smooth variations, where [Mg/Fe] values tend to decrease towards the center of NGC\,3351 and along the direction of the bar. This coupling between the behavior of the [Mg/Fe] and the total metallicity maps indicates a long-lasting chemical enrichment and thus star formation activity in both the nuclear stellar disk and the bar of NGC\,3351.

The bottom right panel of Fig.~\ref{fig:maps} shows the IMF slope map of NGC\,3351. Reassuringly, the measured IMF slope values are close to the Milky Way standard ($\Gamma = 1.3$), with a slight trend following changes in local metallicity. In this regard, the metal-rich stellar population of the bar \citep{Justus2020} are characterized by relatively steeper IMF slopes. This relation between IMF slope and metallicity is similar to what has been already reported in massive ETGs \citep[e.g.][]{MN15c,MN21,Parikh,Sarzi18,Zhou19}. On top of that, the edges of the nuclear star-forming ring seem to exhibit relatively steeper IMF slopes, in particular in the area around the youngest star forming knots. Notice that these bins with a step IMF slope do not correspond one-to-one to the CP regions, suggesting therefore that they are not the result of a trivial systematic effect. Finally, it is worth emphasizing that the steps described above in order to account for an extended star formation history and the presence of young stars, in particular supergiants, are mandatory in order to obtain reliable IMF measurements. To exemplify the importance of these steps, in Appendix~\ref{sec:wrong} we show the (incorrect) IMF slope map resulting from analyzing the spectra of NGC\,3351 following the same approach as for quiescent ETGs.

\begin{figure}
   \centering
   \includegraphics[width=8.5cm]{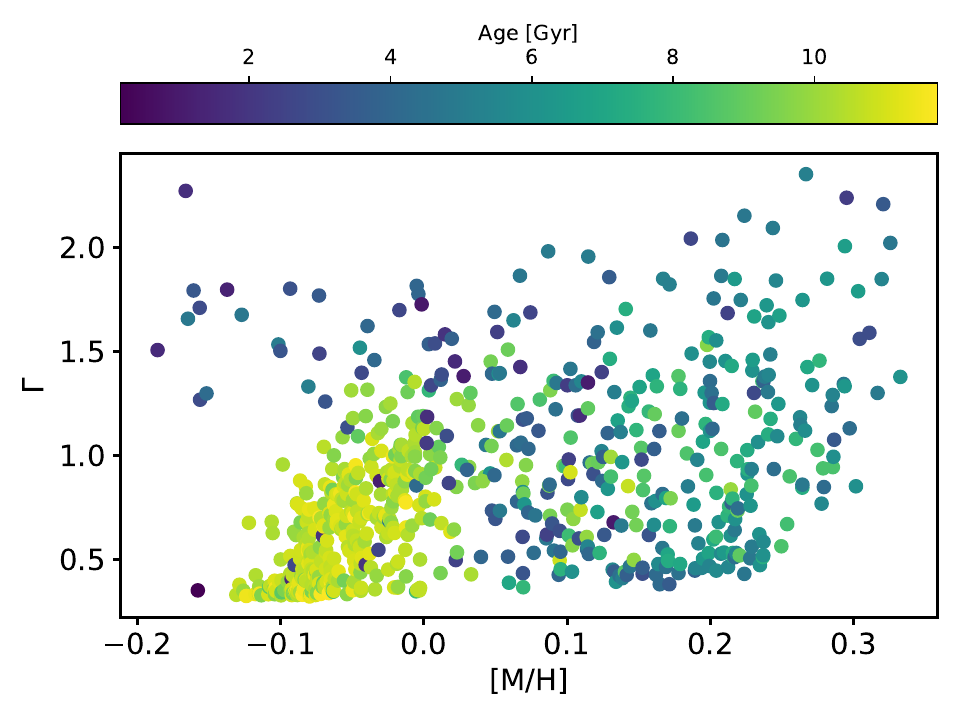}
   \includegraphics[width=8.5cm]{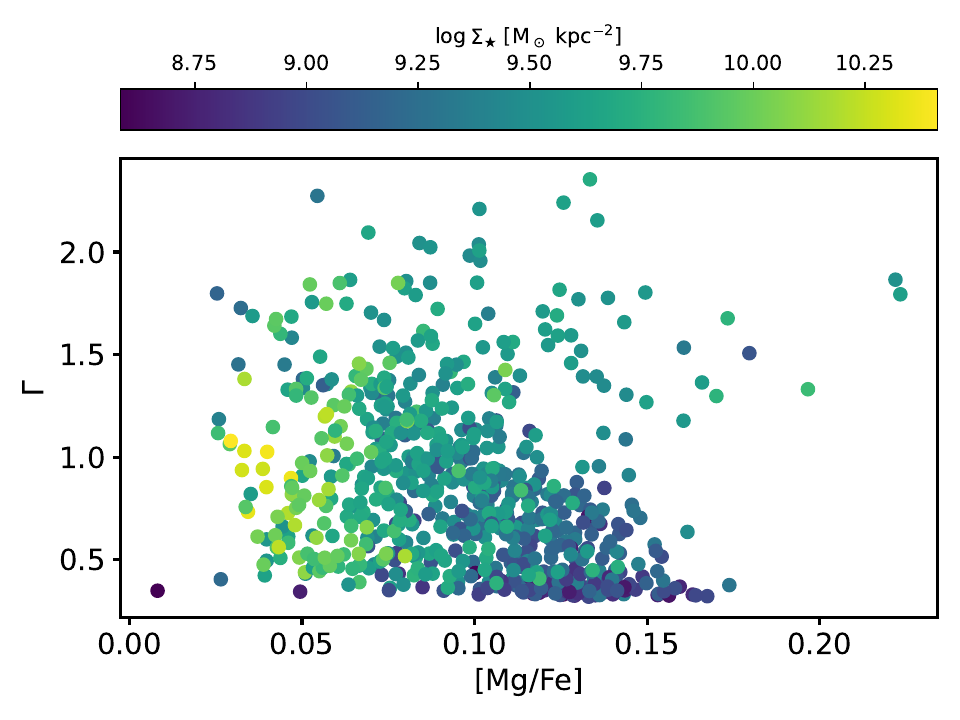}
   \includegraphics[width=8.5cm]{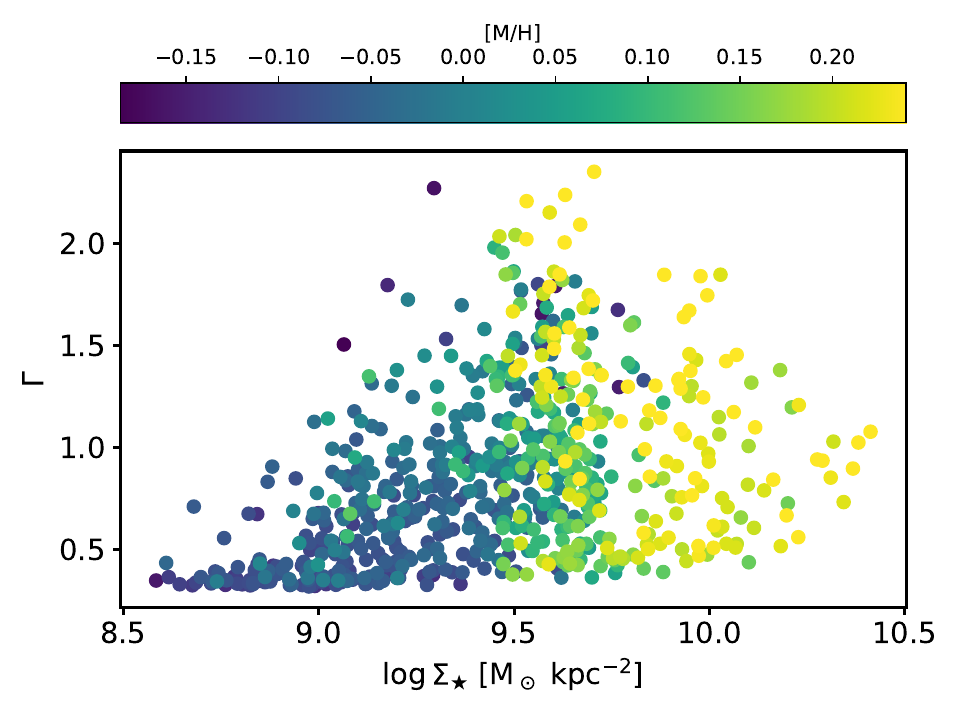}
   \caption{The measured slope of the IMF is shown as a function of metallicity (top panel), [Mg/Fe] (middle panel), and stellar surface mass density (bottom panel). In each panel, symbols have been color coded depending on the age (top), stellar surface mass density (middle) and total metallicity (bottom). Measured IMF values appear to correlate with the three quantities, with the IMF becoming increasingly steeper for more metal-rich, less [Mg/Fe]-enhanced, and denser regions in NGC\,3351.
   }
   \label{fig:trends}
\end{figure}

Figure~\ref{fig:trends} shows how the measured IMF slopes change as a function of metallicity (top panel), [Mg/Fe] (middle panel), and stellar surface mass density (bottom panel). The trend with metallicity already hinted by the maps in Fig.~\ref{fig:maps} becomes evident in the top panel, consistent with IMF measurements in ETGs \citep{MN15b,Parikh,Sarzi18,Zhou19}. Although the scatter increases towards younger ages, IMF variations seem to be dependent on the local metallicity over the whole age range probed in our maps. Moreover, these metal rich and thus chemically evolved regions also tend to exhibit lower [Mg/Fe] values and thus a subtle relation between $\Gamma$ and [Mg/Fe] emerges in the middle panel of Fig.~\ref{fig:trends}. Finally, the measured $\Gamma$ values correlate with the local stellar mass density, which is not surprising given the fact that the evolution of both quantities is expected to be tightly connected \citep[e.g.][]{Sebastianmet,Yulong}. A correlation between stellar mass density and IMF slope has also been reported in ETGs \citep{LB19,vdk17,MN21}.

\section{Discussion} \label{sec:discussion}

\subsection{Is the IMF universal in late-type galaxies?}

\begin{figure*}
   \centering
   \includegraphics[width=14cm]{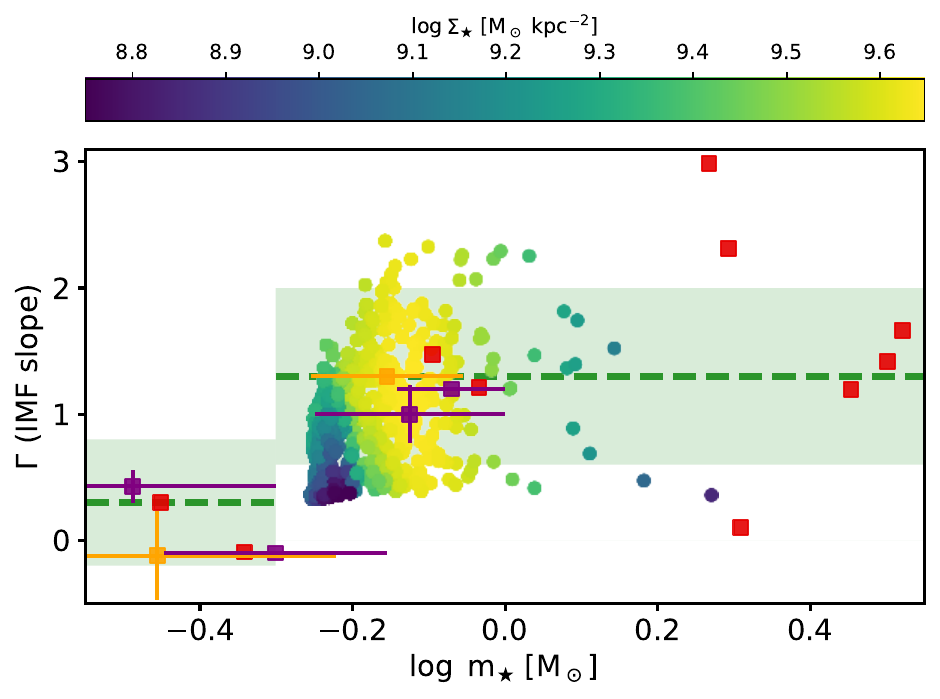}
   \caption{The measured slope of the IMF is shown as a function of the (logarithmic) stellar mass for a compilation of Milky Way measurements (filled squares) and for our measurements of NGC\,3351 (colored circles). Red squares are measurements from \citet{Scalo98}, purple ones are globular cluster measurements from \citet{Piotto99} and finally yellow symbols are IMF measurements of the Milky Way bulge presented in \citet{Holtzman98} and \citet{Zoccali00}. Green dashed lines and shaded areas are the solar neighborhood standard measured by \citet{kroupa}. Our IMF slope measurements, color-coded by the local surface stellar mass density, follow a behavior similar to the resolved IMF studies, although the scatter in our measurements is not purely stochastic and a systematic dependence on the stellar density is found.
   }
   \label{fig:alpha}
\end{figure*}

For quiescent galaxies, with typical ages of around 10 Gyr, only star with masses $\sim 0.5$ M\sun \ contribute to the observed spectra. Therefore, the observational evidence supporting the non-universality of the IMF in these systems only applies to its low-mass end \citep[e.g.][]{Conroy17}. On the other hand, IMF measurements of resolved stellar populations are able to constrain the shape of the IMF across a wide range of stellar masses and this is commonly done by characterizing the slope of the IMF $\Gamma$ as a function of stellar mass. This is illustrated in Fig.~\ref{fig:alpha} where red filled squares are measurements compiled in \citet{Scalo98}, yellow squares are globular clusters from \citet{Piotto99}, and purple squares correspond to the bulge of the Milky Way by \citet{Holtzman98} and \citet{Zoccali00}. Horizontal green dashed lines in Fig.~\ref{fig:alpha} indicate the average IMF slope of the solar neighborhood as measured by \citet{kroupa}, while the shaded green is the associated uncertainty. These measurements in green are commonly adopted as the Milky Way standard. 

Bridging the gap between resolved and unresolved IMF measurements, colored circles in Fig.~\ref{fig:alpha} represent the best-fitting IMF slope of each Voroni bin in NGC\,3351 as a function of the mean stellar mass of the underlying stellar population. Note that the horizontal axis in Fig.~\ref{fig:alpha} does not correspond to the surface stellar mass density of the different bins but to the mean mass of the stars still alive and thus contributing to the observed spectra. In practice, m$_\star$ was calculated as the age-depedent arithmetic mean between the maximum and minimum stellar mass in the BaSTI set of isochrones. For old stellar populations, this mean m$_\star$ is roughly equal to $\sim0.5$\msun, i.e., the mass range explored by IMF studies in quiescent galaxies. The young populations within NGC\,3351 allow us to expand IMF measurements from integrated spectra towards more massive stars. This approximation to the mean stellar mass is independent of the adopted IMF parametrization but it does depend on the exact combination of spectral indices used in the analysis. Attempting a more accurate characterization of the exact mass range probed by each absorption spectra is beyond the scope of this paper but it will be addressed in the future.

The agreement between our IMF measurements and those based on resolved stars in the Milky Way is remarkable. Not only the average value is consistent with observations of the Milky Way but the observed scatter also matches the dispersion around the solar neighborhood. In addition, and mimicking the behavior observed in the Milky Way, we recover flatter IMF slope values when probing lower mass (i.e. older) stars. Note that in our case there is a limit in the minimum stellar mass we can measure with our approach as the average stellar mass of any population older than $\sim$10 Gyr remains essentially constant at $\sim0.5$ M\sun \ \citep[e.g.][]{labarbera}. Yet, and despite of these important methodological differences, our IMF slope measurements closely follow the Milky Way expectations. 

Going beyond the comparison with the Milky Way, the color-coding of our measurements shown in Fig.~\ref{fig:alpha} quantifies the systematic behavior of the IMF map of NGC\,3351, as the measured slope correlates with the surface stellar mass density of each bin. As noted above, IMF variations in NGC\,3351 are not purely stochastic around a mean value of $\Gamma=1.3$ but appear spatially coherent, tracing some of the morphological structures of the galaxy and changes in the stellar population properties (Fig.~\ref{fig:trends}), in agreement with IMF measurements in massive ETGs as noted above \citep[e.g.][]{MN15b,MN21,vdk17,LB19}.

Finally, it is worth reflecting on some of the implicit assumptions and limitations of our approach. In particular, characterizing the slope of the IMF with a single power-law leads to well-known issues with the predicted stellar mass-to-light ratio because of the unrealistically high  contribution of very low-mass stars and stellar remnants to the mass budget $M/L$ \citep[see e.g.][]{ferreras}. Thus, IMF slope measurements presented here should not be na\"ively translated into $M/L$'s. Moreover, the interpretation of the measured $\Gamma$ as a proxy for the \textit{local} shape of the IMF (i.e. for the slope of the IMF over a range of stellar masses) becomes less robust if there is a change in the underlying IMF slope over the explored mass range. This could be particularly problematic for young stellar populations where stars with very different masses contribute to the observed spectrum. To assess the dependence of our results on the adopted IMF slope, we have repeated our stellar population analysis using a broken power law IMF parametrization. The results of this tests are included in Appendix~\ref{sec:bimodal} and are very similar the trends shown in Fig.~\ref{fig:alpha}. Neither of the two IMF functional forms are however realistic enough and more flexible forms should be implemented in the future.

Furthermore, as suggested by Fig.~\ref{fig:alpha}, a change in the age of the observed spectrum can naturally lead to a change in the measured slope of a single power-law IMF just because older populations probe lower stellar masses of the IMF. Despite of these caveats and until newer generations of stellar population models and fitting approaches are available, the adoption of a single power-law IMF parametrization greatly simplifies the analysis and interpretation of the observed spectra. To conclude, our inference relies on the assumption that the IMF is fully sampled. For young stellar populations in spatially resolved nearby galaxies, an incomplete sampling of the IMF could result in important biases. In our case, this would only apply to a handful of Voroni bins and thus our results are robust, but the importance of this effect cannot be overlooked in younger nearby galaxies.

\subsection{The chemical enrichment of the star-forming ring}

As described above, the star-forming ring of NGC\,3351 seems to contain a series of metal-poor regions. \citet{Emsellem22} argued that the chemical enrichment of such low-metallicity regions would be inconsistent with the mixing time-scales of the interstellar medium \citep[e.g.][]{Ho17}, and \citet{Pessa23} performed a thorough analysis of the possible model systematics that could be biasing the fitting of these young stellar populations. Our stellar population approach, based on the analysis of individual absorption features, is different from the full spectral fitting scheme followed by  \citet{Emsellem22}  and \citet{Pessa23} and yet, leads to similar, low-metallicity best-fitting values. 

\begin{figure}
   \centering
   \includegraphics[width=8.5cm]{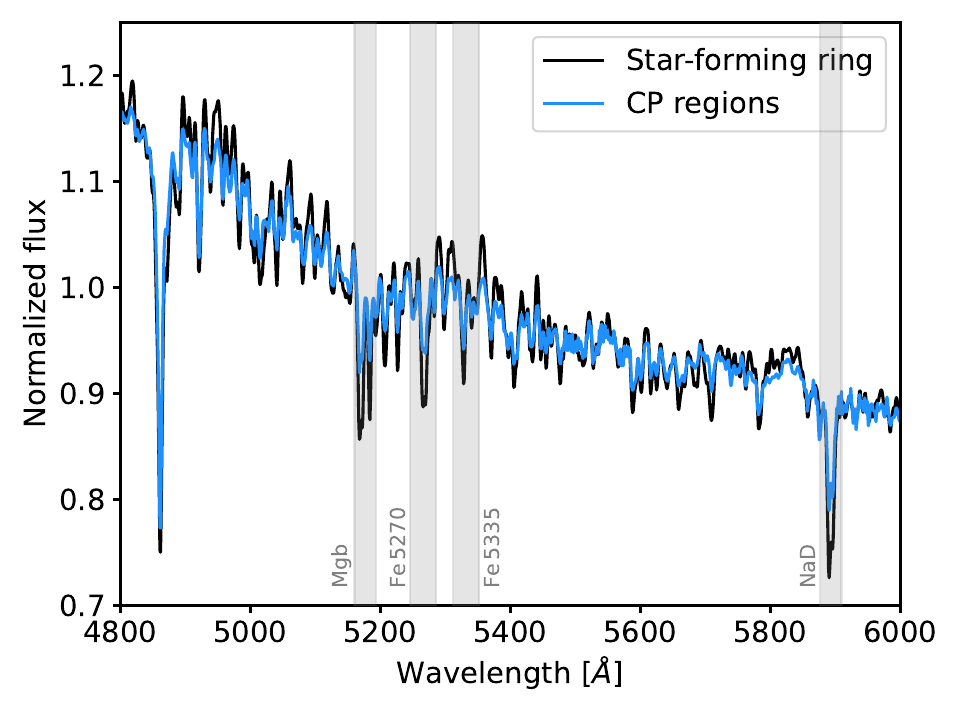}
   \caption{Star-forming ring spectra. The average spectrum of the CP regions of NGC\,3351 is shown in blue, along with the average spectrum of the star-forming ring in black. The depth of the metallicity-sensitive absorption features (indicated with grey shaded regions) is clearly weaker in the CP spectrum. These relatively weak absorption features lead to the observed low-metallicity and high [Mg/Fe] values.
   }
   \label{fig:spec}
\end{figure}

Our extension of the $\alpha$-enhanced MILES models towards young ages allows us to measure, not only the total metallicity content of these regions, but also to be sensitive to variations in the abundance ratio. In \S~\ref{sec:results} we noted that the measured [Mg/Fe] is not the same across all the star-forming and metal-poor regions, as there are four high-[Mg/Fe] knots clearly visible in the map of NGC\,3351. To understand the origin of these four chemically-peculiar (CP) regions, Fig.~\ref{fig:spec} shows the mean optical spectrum of the CP regions (in blue) compared to the average absorption spectrum of the star-forming ring (in black). From this figure it is clear that the CP regions are characterized by weak but clearly-visible metallic absorption features. As discussed in \citet{Pessa23}, it is possible that the presence of a feature-less nebular emission continuum \citep[e.g.][]{Byler17}, associated to the youngest-star forming regions, can dilute the strength of the observed absorption features, which would in turn lead to unrealistically low metallicities. Significant improvements have been in fact achieved over the last years by simultaneously and consistently fitting the stellar absorption and gas emission spectra of galaxies \citep[e.g.][]{fado,Cardoso19,Cardoso22}. We tested this hypothesis by re-running our stellar population analysis but including the presence of a non-stellar continuum as an additional free parameter. As expected, this test resulted in more metal-rich populations once a possible additive continuum is included, but still the recovered metallicity values remain low. Moreover, the presence of such nebular continuum cannot easily explain the relative difference between the strength of Fe and Mg lines (i.e. the apparent [Mg/Fe] enhancement.)

The fact that there is a clear decoupling between the metallicity and the [Mg/Fe] behavior across the star-forming ring of NGC\,3351 opens a suggestive alternative. While magnesium is rapidly produced in core-collapse supernovae, the release of iron-peak elements is delayed as Type Ia supernovae are the main polluters \citep[e.g.][]{Kobayashi20}. The differences between [Mg/Fe] and metallicity maps can be therefore understood as different stages of the chemical evolution process, where the CP regions are just too young for Type Ia supernovae to have enriched the interstellar medium with iron. In this scenario, the metal-poor but [Mg/Fe]$\sim0$ knots would be more chemically-evolved regions, although still powered by relatively metal-poor gas \citep[e.g.][]{Maiolino19}. 

The comparison with the photometric analysis of \citet{Calzetti21} sheds further light on the chemical enrichment of the star-forming ring of of NGC\,3351. The four CP visible in our [Mg/Fe] plane correspond exactly with those regions where the multi-band photometric analysis reveals the presence of very young ($\lesssim 10$ Myr) star-formation bursts, too young actually for Type Ia supernovae to have exploded. Complementary, our metal-poor regions with solar [Mg/Fe] values are located in regions with a more extended star formation history over the last $\sim300$ Myr. These differences between the photometric star-formation time-scales of the different regions, although calculated completely independently, coincide with the chemical time-scales derived from our stellar population analysis. 

Finally, it is worth noticing that, although the luminosity-weighted age of the CP regions is slightly older than the age of the metal-poor and [Mg/Fe]$\sim0$ knots (see \S~\ref{sec:results}), that does not invalidate the proposed scenario. On the contrary, according to our measured star formation histories, up to 55\% of the flux of the CP regions is contributed by stars younger than 10 Myr, while for the metal-poor and [Mg/Fe]$\sim0$ that percentage goes down to 35\%. It is precisely the extended star formation history of these long-lasting bursts what leads to, on average, younger mean ages. Our results are in notable agreement with the findings of \citet{Calzetti21} and highlight the limitations of using mean values to characterize the age of stellar populations with complex and extended star formation histories.

As a cautionary note, however, these results regarding the potential chemical peculiarities of the star-forming knots of NGC\,3351, although show encouraging consistency with the independent analysis of \citet{Calzetti21}, are sensitive to significant model systematics and limitations. In particular, the sensitivity of integrated spectra to variations in the chemical composition, and in particular to the abundance pattern of stars, becomes progressively weaker in younger stellar populations. Moreover, from a modelling perspective, the theoretical corrections of \citet{Coelho05} applied to the \citet{Vazdekis15} SSP models are limited to stellar atmospheres cooler than 7000 K, which is insufficient to properly cover the whole temperature range explored in this work. A more reliable modelling of the chemical composition of young stellar populations \citep[e.g.][]{Percival09} stands therefore a natural step forward for upcoming works, combined with the analysis of the gas phase chemistry \citep[e.g.][]{Diaz07,Moustakas10}.

\section{Summary and conclusions} \label{sec:summary}

Making use of the tools and ideas that have revealed the non-universality of the IMF in massive ETGs, we have developed a stellar population fitting approach capable of analyzing the absorption spectra of galaxies with young populations and complex star formation histories. We have measured the spatially-resolved age, metallicity, [Mg/Fe], and IMF slope $\Gamma$ maps of the late-type galaxy NGC\,3351, similar to the Milky Way in mass and characterized by the presence of a nuclear disk and star-forming ring. Our main results can be summarized as follows:

\begin{itemize}
   \item The nuclear and morphological structures of NGC\,3351 are clearly visible on its stellar population maps. The nuclear disk stands as a relatively young and metal-rich component, with a [Mg/Fe]$\sim$0. Similarly, the bar of NGC\,3351 hosts metal-enriched and low [Mg/Fe] stars, a clear indication of chemically-evolved populations. 
   
   \item  The measured values of the IMF slope in NGC\,3351 are consistent with an underlying Milky Way-like IMF. However, embedded in this \textit{universal} behavior, the slope of the IMF changes in a systematic and spatially-coherent manner. In particular, the bar seems to be characterized by a relatively steep IMF slope compared to the surrounding inner parts of the disk. Moreover, the outskirts of the star-forming nuclear ring show evidence of steep IMF values as well. 
   
   \item Regions with peculiar chemical properties are measured within the nuclear star-forming ring. These regions, associated with star-forming bursts, are characterized by the presence of metal-poor populations. Interestingly, the measured [Mg/Fe] values over these regions show a bimodal distribution, with either [Mg/Fe]$\sim0$ or [Mg/Fe]$\sim0.3$. We interpret this decoupling between total metallicity and [Mg/Fe] as the chemical signature of core-collapse supernova enrichment, in agreement with the (independent) spectroscopic and photometric age determinations.
\end{itemize}

In summary, our results highlight the feasibility of measuring in detail the properties of young stellar populations based on their integrated absorption spectra. When analyzed in the context of understanding the origin of the observed variations in the IMF, these young populations are a natural bridge between the universality of the Milky Way and the systematic changes measured in the centers of massive ETGs. In this regard, the IMF map of NGC\,3351 allows for a fine and complete characterization of the IMF across the inner regions of the galaxy, contrary to the relatively scarce resolved IMF measurements. It is only because of our large number of independent IMF measurements that we are able to detect the subtle IMF variations that appear to be mostly related to its low-mass end. Although observations in the Milky Way seem to favor some sort of universal IMF shape, it is possible that our ability to detect systematic deviations is hampered by the limited number of measurements. 

Looking ahead, the detailed modelling of the absorption spectra of young stellar populations will play a capital role in understanding the high redshift Universe in the JWST and ELT era. In order to maximize the amount of information extracted from these observations, models and analysis tools have to become flexible enough to account for the complex and potentially unfamiliar behavior of the young Universe. The methodology presented here is a step in that direction but several aspects need to be improved, from a better treatment of non-canonical ingredients to a self-consistent measurement of the time evolution of the stellar population properties which takes into account the potentially time-varying nature of the IMF.

\begin{acknowledgement}
We would like to thank the referee for providing insightful and relevant comments that significantly improved our manuscript. Raw and reduced data are available at the ESO Science Archive Facility. IMM would like to thank Alexandre Vazdekis and Alina B\"ocker for their insightful discussions during the preparation of this manuscript. We acknowledge support from grants PID2019-107427GB-C32,  PID2019-107427GB-C31 from the Spanish Ministry of Science and Innovation. This work was supported by STFC grants ST/T000244/1 and ST/X001075/1, based on observations collected at the European Southern Observatory under ESO programme 097.B-0640(A). J.M.A. acknowledges the support of the Viera y Clavijo Senior program funded by ACIISI and ULL. J.M.A. and A.dLC. acknowledges support from the Agencia Estatal de Investigaci\'on del Ministerio de Ciencia e Innovaci\'on (MCIN/AEI/ 10.13039/501100011033) under grant (PID2021-128131NB-I00) and the European Regional Development Fund (ERDF) "A way of making europe". JN acknowledges funding from the European Research Council (ERC) under the European Union’s Horizon 2020 research and innovation programme (grant agreement No. 694343). IP acknowledges financial support by the research project
PID2020-113689GB-I00 financed by MCIN/AEI/10.13039/501100011033
\end{acknowledgement}

\bibliographystyle{aa}  

\begin{thebibliography}{155}
   \expandafter\ifx\csname natexlab\endcsname\relax\def\natexlab#1{#1}\fi
   
   \bibitem[{{Asa'd} {et~al.}(2017){Asa'd}, {Vazdekis}, {Cervi{\~n}o}, {No{\"e}l},
     {Beasley}, \& {Kassab}}]{youngMILES}
   {Asa'd}, R.~S., {Vazdekis}, A., {Cervi{\~n}o}, M., {et~al.} 2017, \mnras, 471,
     3599
   
   \bibitem[{{Athanassoula}(1992)}]{Lia92}
   {Athanassoula}, E. 1992, \mnras, 259, 345
   
   \bibitem[{{Auger} {et~al.}(2010){Auger}, {Treu}, {Gavazzi}, {Bolton},
     {Koopmans}, \& {Marshall}}]{auger}
   {Auger}, M.~W., {Treu}, T., {Gavazzi}, R., {et~al.} 2010, \apjl, 721, L163
   
   \bibitem[{{Bacon} {et~al.}(2010){Bacon}, {Accardo}, {Adjali}, {Anwand},
     {Bauer}, {Biswas}, {Blaizot}, {Boudon}, {Brau-Nogue}, {Brinchmann},
     {Caillier}, {Capoani}, {Carollo}, {Contini}, {Couderc}, {Daguis{\'e}},
     {Deiries}, {Delabre}, {Dreizler}, {Dubois}, {Dupieux}, {Dupuy}, {Emsellem},
     {Fechner}, {Fleischmann}, {Fran{\c c}ois}, {Gallou}, {Gharsa}, {Glindemann},
     {Gojak}, {Guiderdoni}, {Hansali}, {Hahn}, {Jarno}, {Kelz}, {Koehler},
     {Kosmalski}, {Laurent}, {Le Floch}, {Lilly}, {Lizon}, {Loupias}, {Manescau},
     {Monstein}, {Nicklas}, {Olaya}, {Pares}, {Pasquini}, {P{\'e}contal-Rousset},
     {Pell{\'o}}, {Petit}, {Popow}, {Reiss}, {Remillieux}, {Renault}, {Roth},
     {Rupprecht}, {Serre}, {Schaye}, {Soucail}, {Steinmetz}, {Streicher}, {Stuik},
     {Valentin}, {Vernet}, {Weilbacher}, {Wisotzki}, \& {Yerle}}]{Bacon10}
   {Bacon}, R., {Accardo}, M., {Adjali}, L., {et~al.} 2010, in \procspie, Vol.
     7735, Ground-based and Airborne Instrumentation for Astronomy III, 773508
   
   \bibitem[{{Barbosa} {et~al.}(2021){Barbosa}, {Spiniello}, {Arnaboldi},
     {Coccato}, {Hilker}, \& {Richtler}}]{Barbosa20}
   {Barbosa}, C.~E., {Spiniello}, C., {Arnaboldi}, M., {et~al.} 2021, \aap, 649,
     A93
   
   \bibitem[{{Bastian} {et~al.}(2010){Bastian}, {Covey}, \& {Meyer}}]{bastian}
   {Bastian}, N., {Covey}, K.~R., \& {Meyer}, M.~R. 2010, \araa, 48, 339
   
   \bibitem[{{Bertelli} {et~al.}(1994){Bertelli}, {Bressan}, {Chiosi}, {Fagotto},
     \& {Nasi}}]{Padova94}
   {Bertelli}, G., {Bressan}, A., {Chiosi}, C., {Fagotto}, F., \& {Nasi}, E. 1994,
     \aaps, 106, 275
   
   \bibitem[{{Bittner} {et~al.}(2021){Bittner}, {de Lorenzo-C{\'a}ceres},
     {Gadotti}, {S{\'a}nchez-Bl{\'a}zquez}, {Neumann}, {Coelho},
     {Falc{\'o}n-Barroso}, {Fragkoudi}, {Kim}, {Mart{\'\i}n-Navarro},
     {M{\'e}ndez-Abreu}, {P{\'e}rez}, {Querejeta}, \& {van de Ven}}]{Bittner21}
   {Bittner}, A., {de Lorenzo-C{\'a}ceres}, A., {Gadotti}, D.~A., {et~al.} 2021,
     \aap, 646, A42
   
   \bibitem[{{Bittner} {et~al.}(2020){Bittner}, {S{\'a}nchez-Bl{\'a}zquez},
     {Gadotti}, {Neumann}, {Fragkoudi}, {Coelho}, {de Lorenzo-C{\'a}ceres},
     {Falc{\'o}n-Barroso}, {Kim}, {Leaman}, {Mart{\'\i}n-Navarro},
     {M{\'e}ndez-Abreu}, {P{\'e}rez}, {Querejeta}, {Seidel}, \& {van de
     Ven}}]{Bittner20}
   {Bittner}, A., {S{\'a}nchez-Bl{\'a}zquez}, P., {Gadotti}, D.~A., {et~al.} 2020,
     \aap, 643, A65
   
   \bibitem[{{Brown} \& {Wilson}(2019)}]{Brown19}
   {Brown}, T. \& {Wilson}, C.~D. 2019, \apj, 879, 17
   
   \bibitem[{{Byler} {et~al.}(2017){Byler}, {Dalcanton}, {Conroy}, \&
     {Johnson}}]{Byler17}
   {Byler}, N., {Dalcanton}, J.~J., {Conroy}, C., \& {Johnson}, B.~D. 2017, \apj,
     840, 44
   
   \bibitem[{{Calzetti} {et~al.}(2021){Calzetti}, {Battisti}, {Shivaei}, {Messa},
     {Cignoni}, {Adamo}, {Dale}, {Gallagher}, {Grasha}, {Grebel}, {Kennicutt},
     {Linden}, {{\"O}stlin}, {Sabbi}, {Smith}, {Tosi}, \& {Wofford}}]{Calzetti21}
   {Calzetti}, D., {Battisti}, A.~J., {Shivaei}, I., {et~al.} 2021, \apj, 913, 37
   
   \bibitem[{{Cappellari}(2008)}]{Cappellari08}
   {Cappellari}, M. 2008, \mnras, 390, 71
   
   \bibitem[{{Cappellari}(2017)}]{Cappellari17}
   {Cappellari}, M. 2017, \mnras, 466, 798
   
   \bibitem[{{Cappellari} \& {Copin}(2003)}]{voronoi}
   {Cappellari}, M. \& {Copin}, Y. 2003, \mnras, 342, 345
   
   \bibitem[{{Cappellari} \& {Emsellem}(2004)}]{ppxf}
   {Cappellari}, M. \& {Emsellem}, E. 2004, \pasp, 116, 138
   
   \bibitem[{{Cappellari} {et~al.}(2012){Cappellari}, {McDermid}, {Alatalo},
     {Blitz}, {Bois}, {Bournaud}, {Bureau}, {Crocker}, {Davies}, {Davis}, {de
     Zeeuw}, {Duc}, {Emsellem}, {Khochfar}, {Krajnovi{\'c}}, {Kuntschner},
     {Lablanche}, {Morganti}, {Naab}, {Oosterloo}, {Sarzi}, {Scott}, {Serra},
     {Weijmans}, \& {Young}}]{cappellari}
   {Cappellari}, M., {McDermid}, R.~M., {Alatalo}, K., {et~al.} 2012, \nat, 484,
     485
   
   \bibitem[{{Cappellari} {et~al.}(2013){Cappellari}, {McDermid}, {Alatalo},
     {Blitz}, {Bois}, {Bournaud}, {Bureau}, {Crocker}, {Davies}, {Davis}, {de
     Zeeuw}, {Duc}, {Emsellem}, {Khochfar}, {Krajnovi{\'c}}, {Kuntschner},
     {Morganti}, {Naab}, {Oosterloo}, {Sarzi}, {Scott}, {Serra}, {Weijmans}, \&
     {Young}}]{Cappellari13}
   {Cappellari}, M., {McDermid}, R.~M., {Alatalo}, K., {et~al.} 2013, \mnras, 432,
     1862
   
   \bibitem[{{Cardoso} {et~al.}(2019){Cardoso}, {Gomes}, \&
     {Papaderos}}]{Cardoso19}
   {Cardoso}, L. S.~M., {Gomes}, J.~M., \& {Papaderos}, P. 2019, \aap, 622, A56
   
   \bibitem[{{Cardoso} {et~al.}(2022){Cardoso}, {Gomes}, {Papaderos},
     {Pappalardo}, {Miranda}, {Paulino-Afonso}, {Afonso}, \& {Lagos}}]{Cardoso22}
   {Cardoso}, L. S.~M., {Gomes}, J.~M., {Papaderos}, P., {et~al.} 2022, \aap, 667,
     A11
   
   \bibitem[{{Cervantes} \& {Vazdekis}(2009)}]{Cervantes}
   {Cervantes}, J.~L. \& {Vazdekis}, A. 2009, \mnras, 392, 691
   
   \bibitem[{{Chabrier}(2003)}]{Chabrier}
   {Chabrier}, G. 2003, \pasp, 115, 763
   
   \bibitem[{{Chabrier} {et~al.}(2014){Chabrier}, {Hennebelle}, \&
     {Charlot}}]{Chabrier14}
   {Chabrier}, G., {Hennebelle}, P., \& {Charlot}, S. 2014, \apj, 796, 75
   
   \bibitem[{{Cid Fernandes} {et~al.}(2005){Cid Fernandes}, {Mateus}, {Sodr{\'e}},
     {Stasi{\'n}ska}, \& {Gomes}}]{CF05}
   {Cid Fernandes}, R., {Mateus}, A., {Sodr{\'e}}, L., {Stasi{\'n}ska}, G., \&
     {Gomes}, J.~M. 2005, \mnras, 358, 363
   
   \bibitem[{{Coelho} {et~al.}(2005){Coelho}, {Barbuy}, {Mel{\'e}ndez},
     {Schiavon}, \& {Castilho}}]{Coelho05}
   {Coelho}, P., {Barbuy}, B., {Mel{\'e}ndez}, J., {Schiavon}, R.~P., \&
     {Castilho}, B.~V. 2005, \aap, 443, 735
   
   \bibitem[{{Conroy} {et~al.}(2014){Conroy}, {Graves}, \& {van
     Dokkum}}]{Conroy14}
   {Conroy}, C., {Graves}, G.~J., \& {van Dokkum}, P.~G. 2014, \apj, 780, 33
   
   \bibitem[{{Conroy} \& {van Dokkum}(2012{\natexlab{a}})}]{conroy}
   {Conroy}, C. \& {van Dokkum}, P. 2012{\natexlab{a}}, \apj, 747, 69
   
   \bibitem[{{Conroy} \& {van Dokkum}(2012{\natexlab{b}})}]{conroy12}
   {Conroy}, C. \& {van Dokkum}, P.~G. 2012{\natexlab{b}}, \apj, 760, 71
   
   \bibitem[{{Conroy} {et~al.}(2017){Conroy}, {van Dokkum}, \&
     {Villaume}}]{Conroy17}
   {Conroy}, C., {van Dokkum}, P.~G., \& {Villaume}, A. 2017, \apj, 837, 166
   
   \bibitem[{{Corsini} {et~al.}(2017){Corsini}, {Wegner}, {Thomas}, {Saglia}, \&
     {Bender}}]{Corsini17}
   {Corsini}, E.~M., {Wegner}, G.~A., {Thomas}, J., {Saglia}, R.~P., \& {Bender},
     R. 2017, \mnras, 466, 974
   
   \bibitem[{{da Cunha} {et~al.}(2008){da Cunha}, {Charlot}, \& {Elbaz}}]{Cunha}
   {da Cunha}, E., {Charlot}, S., \& {Elbaz}, D. 2008, \mnras, 388, 1595
   
   \bibitem[{{Davis} \& {McDermid}(2017)}]{Davis17}
   {Davis}, T.~A. \& {McDermid}, R.~M. 2017, \mnras, 464, 453
   
   \bibitem[{{Davis} \& {van de Voort}(2020)}]{Davis20}
   {Davis}, T.~A. \& {van de Voort}, F. 2020, \mnras, 498, 4051
   
   \bibitem[{{de Lorenzo-C{\'a}ceres} {et~al.}(2019){de Lorenzo-C{\'a}ceres},
     {S{\'a}nchez-Bl{\'a}zquez}, {M{\'e}ndez-Abreu}, {Gadotti},
     {Falc{\'o}n-Barroso}, {Mart{\'\i}nez-Valpuesta}, {Coelho}, {Fragkoudi},
     {Husemann}, {Leaman}, {P{\'e}rez}, {Querejeta}, {Seidel}, \& {van de
     Ven}}]{Adri19}
   {de Lorenzo-C{\'a}ceres}, A., {S{\'a}nchez-Bl{\'a}zquez}, P.,
     {M{\'e}ndez-Abreu}, J., {et~al.} 2019, \mnras, 484, 5296
   
   \bibitem[{{D{\'\i}az} {et~al.}(2007){D{\'\i}az}, {Terlevich}, {Castellanos}, \&
     {H{\"a}gele}}]{Diaz07}
   {D{\'\i}az}, {\'A}.~I., {Terlevich}, E., {Castellanos}, M., \& {H{\"a}gele},
     G.~F. 2007, \mnras, 382, 251
   
   \bibitem[{{Dutton} {et~al.}(2012){Dutton}, {Mendel}, \& {Simard}}]{Dutton12}
   {Dutton}, A.~A., {Mendel}, J.~T., \& {Simard}, L. 2012, \mnras, 422, L33
   
   \bibitem[{{Eldridge} {et~al.}(2017){Eldridge}, {Stanway}, {Xiao}, {McClelland},
     {Taylor}, {Ng}, {Greis}, \& {Bray}}]{Eldridge17}
   {Eldridge}, J.~J., {Stanway}, E.~R., {Xiao}, L., {et~al.} 2017, \pasa, 34, e058
   
   \bibitem[{{Emsellem} {et~al.}(2015){Emsellem}, {Renaud}, {Bournaud},
     {Elmegreen}, {Combes}, \& {Gabor}}]{Emsellem15}
   {Emsellem}, E., {Renaud}, F., {Bournaud}, F., {et~al.} 2015, \mnras, 446, 2468
   
   \bibitem[{{Emsellem} {et~al.}(2022){Emsellem}, {Schinnerer}, {Santoro},
     {Belfiore}, {Pessa}, {McElroy}, {Blanc}, {Congiu}, {Groves}, {Ho}, {Kreckel},
     {Razza}, {Sanchez-Blazquez}, {Egorov}, {Faesi}, {Klessen}, {Leroy}, {Meidt},
     {Querejeta}, {Rosolowsky}, {Scheuermann}, {Anand}, {Barnes},
     {Be{\v{s}}li{\'c}}, {Bigiel}, {Boquien}, {Cao}, {Chevance}, {Dale},
     {Eibensteiner}, {Glover}, {Grasha}, {Henshaw}, {Hughes}, {Koch}, {Kruijssen},
     {Lee}, {Liu}, {Pan}, {Pety}, {Saito}, {Sandstrom}, {Schruba}, {Sun},
     {Thilker}, {Usero}, {Watkins}, \& {Williams}}]{Emsellem22}
   {Emsellem}, E., {Schinnerer}, E., {Santoro}, F., {et~al.} 2022, \aap, 659, A191
   
   \bibitem[{{Faber}(1973)}]{Faber73}
   {Faber}, S.~M. 1973, \apj, 179, 731
   
   \bibitem[{{Ferreras} {et~al.}(2013){Ferreras}, {La Barbera}, {de la Rosa},
     {Vazdekis}, {de Carvalho}, {Falc{\'o}n-Barroso}, \&
     {Ricciardelli}}]{ferreras}
   {Ferreras}, I., {La Barbera}, F., {de la Rosa}, I.~G., {et~al.} 2013, \mnras,
     429, L15
   
   \bibitem[{{Ferreras} {et~al.}(2022){Ferreras}, {Lahav}, {Somerville}, \&
     {Silk}}]{Ferreras22}
   {Ferreras}, I., {Lahav}, O., {Somerville}, R.~S., \& {Silk}, J. 2022, arXiv
     e-prints, arXiv:2208.05489
   
   \bibitem[{{Ferreras} {et~al.}(2015){Ferreras}, {Weidner}, {Vazdekis}, \& {La
     Barbera}}]{Ferreras15}
   {Ferreras}, I., {Weidner}, C., {Vazdekis}, A., \& {La Barbera}, F. 2015,
     \mnras, 448, L82
   
   \bibitem[{{Fontanot} {et~al.}(2018){Fontanot}, {La Barbera}, {De Lucia},
     {Pasquali}, \& {Vazdekis}}]{Fontanot18}
   {Fontanot}, F., {La Barbera}, F., {De Lucia}, G., {Pasquali}, A., \&
     {Vazdekis}, A. 2018, \mnras, 479, 5678
   
   \bibitem[{{Foreman-Mackey} {et~al.}(2013){Foreman-Mackey}, {Hogg}, {Lang}, \&
     {Goodman}}]{emcee}
   {Foreman-Mackey}, D., {Hogg}, D.~W., {Lang}, D., \& {Goodman}, J. 2013, \pasp,
     125, 306
   
   \bibitem[{{Gadotti} {et~al.}(2020){Gadotti}, {Bittner}, {Falc{\'o}n-Barroso},
     {M{\'e}ndez-Abreu}, {Kim}, {Fragkoudi}, {de Lorenzo-C{\'a}ceres}, {Leaman},
     {Neumann}, {Querejeta}, {S{\'a}nchez-Bl{\'a}zquez}, {Martig},
     {Mart{\'\i}n-Navarro}, {P{\'e}rez}, {Seidel}, \& {van de Ven}}]{Gadotti20}
   {Gadotti}, D.~A., {Bittner}, A., {Falc{\'o}n-Barroso}, J., {et~al.} 2020, \aap,
     643, A14
   
   \bibitem[{{Gadotti} {et~al.}(2019){Gadotti}, {S{\'a}nchez-Bl{\'a}zquez},
     {Falc{\'o}n-Barroso}, {Husemann}, {Seidel}, {P{\'e}rez}, {de
     Lorenzo-C{\'a}ceres}, {Martinez-Valpuesta}, {Fragkoudi}, {Leung}, {van de
     Ven}, {Leaman}, {Coelho}, {Martig}, {Kim}, {Neumann}, \& {Querejeta}}]{TIMER}
   {Gadotti}, D.~A., {S{\'a}nchez-Bl{\'a}zquez}, P., {Falc{\'o}n-Barroso}, J.,
     {et~al.} 2019, \mnras, 482, 506
   
   \bibitem[{{Gallazzi} {et~al.}(2005){Gallazzi}, {Charlot}, {Brinchmann},
     {White}, \& {Tremonti}}]{Gallazzi05}
   {Gallazzi}, A., {Charlot}, S., {Brinchmann}, J., {White}, S. D.~M., \&
     {Tremonti}, C.~A. 2005, \mnras, 362, 41
   
   \bibitem[{{Garc{\'\i}a P{\'e}rez} {et~al.}(2021){Garc{\'\i}a P{\'e}rez},
     {S{\'a}nchez-Bl{\'a}zquez}, {Vazdekis}, {Allende Prieto}, {Milone}, {Sansom},
     {Gorgas}, {Falc{\'o}n-Barroso}, {Mart{\'\i}n Navarro}, \& {Cacho}}]{MILES21}
   {Garc{\'\i}a P{\'e}rez}, A.~E., {S{\'a}nchez-Bl{\'a}zquez}, P., {Vazdekis}, A.,
     {et~al.} 2021, \mnras, 505, 4496
   
   \bibitem[{{Gomes} \& {Papaderos}(2017)}]{fado}
   {Gomes}, J.~M. \& {Papaderos}, P. 2017, \aap, 603, A63
   
   \bibitem[{{Guszejnov} {et~al.}(2019){Guszejnov}, {Hopkins}, \&
     {Graus}}]{Guszejnov19}
   {Guszejnov}, D., {Hopkins}, P.~F., \& {Graus}, A.~S. 2019, \mnras, 485, 4852
   
   \bibitem[{{Han} \& {Han}(2019)}]{Han}
   {Han}, Y. \& {Han}, Z. 2019, \apjs, 240, 3
   
   \bibitem[{{Hennebelle} \& {Chabrier}(2008)}]{Hennebelle08}
   {Hennebelle}, P. \& {Chabrier}, G. 2008, \apj, 684, 395
   
   \bibitem[{{Ho} {et~al.}(2017){Ho}, {Seibert}, {Meidt}, {Kudritzki},
     {Kobayashi}, {Groves}, {Kewley}, {Madore}, {Rich}, {Schinnerer},
     {D'Agostino}, \& {Poetrodjojo}}]{Ho17}
   {Ho}, I.~T., {Seibert}, M., {Meidt}, S.~E., {et~al.} 2017, \apj, 846, 39
   
   \bibitem[{{Holtzman} {et~al.}(1998){Holtzman}, {Watson}, {Baum}, {Grillmair},
     {Groth}, {Light}, {Lynds}, \& {O'Neil}}]{Holtzman98}
   {Holtzman}, J.~A., {Watson}, A.~M., {Baum}, W.~A., {et~al.} 1998, \aj, 115,
     1946
   
   \bibitem[{{Hopkins}(2012)}]{Hopkins12IMF}
   {Hopkins}, P.~F. 2012, \mnras, 423, 2037
   
   \bibitem[{{Hoversten} \& {Glazebrook}(2008)}]{Hoversten08}
   {Hoversten}, E.~A. \& {Glazebrook}, K. 2008, \apj, 675, 163
   
   \bibitem[{{Jeans}(1922)}]{Jeans22}
   {Jeans}, J.~H. 1922, \mnras, 82, 122
   
   \bibitem[{{Johansson} {et~al.}(2012){Johansson}, {Thomas}, \&
     {Maraston}}]{johansson12}
   {Johansson}, J., {Thomas}, D., \& {Maraston}, C. 2012, \mnras, 421, 1908
   
   \bibitem[{{Johnson} {et~al.}(2021){Johnson}, {Leja}, {Conroy}, \&
     {Speagle}}]{prospector}
   {Johnson}, B.~D., {Leja}, J., {Conroy}, C., \& {Speagle}, J.~S. 2021, \apjs,
     254, 22
   
   \bibitem[{{Kim} \& {Stone}(2012)}]{Kim12}
   {Kim}, W.-T. \& {Stone}, J.~M. 2012, \apj, 751, 124
   
   \bibitem[{{Kobayashi} {et~al.}(2020){Kobayashi}, {Karakas}, \&
     {Lugaro}}]{Kobayashi20}
   {Kobayashi}, C., {Karakas}, A.~I., \& {Lugaro}, M. 2020, \apj, 900, 179
   
   \bibitem[{{Kriek} {et~al.}(2009){Kriek}, {van Dokkum}, {Labb{\'e}}, {Franx},
     {Illingworth}, {Marchesini}, \& {Quadri}}]{Kriek09}
   {Kriek}, M., {van Dokkum}, P.~G., {Labb{\'e}}, I., {et~al.} 2009, \apj, 700,
     221
   
   \bibitem[{{Kroupa}(2001)}]{MW}
   {Kroupa}, P. 2001, \mnras, 322, 231
   
   \bibitem[{{Kroupa}(2002)}]{kroupa}
   {Kroupa}, P. 2002, Science, 295, 82
   
   \bibitem[{{Kroupa} \& {Jerabkova}(2019)}]{Kroupa19}
   {Kroupa}, P. \& {Jerabkova}, T. 2019, Nature Astronomy, 3, 482
   
   \bibitem[{{Krumholz}(2011)}]{Krumholz11}
   {Krumholz}, M.~R. 2011, \apj, 743, 110
   
   \bibitem[{{Krumholz}(2014)}]{Krumholz14}
   {Krumholz}, M.~R. 2014, \physrep, 539, 49
   
   \bibitem[{{La Barbera} {et~al.}(2013){La Barbera}, {Ferreras}, {Vazdekis}, {de
     la Rosa}, {de Carvalho}, {Trevisan}, {Falc{\'o}n-Barroso}, \&
     {Ricciardelli}}]{labarbera}
   {La Barbera}, F., {Ferreras}, I., {Vazdekis}, A., {et~al.} 2013, \mnras, 433,
     3017
   
   \bibitem[{{La Barbera} {et~al.}(2019){La Barbera}, {Vazdekis}, {Ferreras},
     {Pasquali}, {Allende Prieto}, {Mart{\'\i}n-Navarro}, {Aguado}, {de Carvalho},
     {Rembold}, {Falc{\'o}n-Barroso}, \& {van de Ven}}]{LB19}
   {La Barbera}, F., {Vazdekis}, A., {Ferreras}, I., {et~al.} 2019, \mnras, 489,
     4090
   
   \bibitem[{{La Barbera} {et~al.}(2016){La Barbera}, {Vazdekis}, {Ferreras},
     {Pasquali}, {Cappellari}, {Mart{\'{\i}}n-Navarro}, {Sch{\"o}nebeck}, \&
     {Falc{\'o}n-Barroso}}]{LB16}
   {La Barbera}, F., {Vazdekis}, A., {Ferreras}, I., {et~al.} 2016, \mnras, 457,
     1468
   
   \bibitem[{{Lagattuta} {et~al.}(2017){Lagattuta}, {Mould}, {Forbes}, {Monson},
     {Pastorello}, \& {Persson}}]{Lagattuta17}
   {Lagattuta}, D.~J., {Mould}, J.~R., {Forbes}, D.~A., {et~al.} 2017, \apj, 846,
     166
   
   \bibitem[{{L{\"a}sker} {et~al.}(2013){L{\"a}sker}, {van den Bosch}, {van de
     Ven}, {Ferreras}, {La Barbera}, {Vazdekis}, \&
     {Falc{\'o}n-Barroso}}]{Lasker13}
   {L{\"a}sker}, R., {van den Bosch}, R.~C.~E., {van de Ven}, G., {et~al.} 2013,
     \mnras, 434, L31
   
   \bibitem[{{Leaman} {et~al.}(2019){Leaman}, {Fragkoudi}, {Querejeta}, {Leung},
     {Gadotti}, {Husemann}, {Falc{\'o}n-Barroso}, {S{\'a}nchez-Bl{\'a}zquez}, {van
     de Ven}, {Kim}, {Coelho}, {Lyubenova}, {de Lorenzo-C{\'a}ceres}, {Martig},
     {Martinez-Valpuesta}, {Neumann}, {P{\'e}rez}, \& {Seidel}}]{Leaman19}
   {Leaman}, R., {Fragkoudi}, F., {Querejeta}, M., {et~al.} 2019, \mnras, 488,
     3904
   
   \bibitem[{{Lee} {et~al.}(2009){Lee}, {Gil de Paz}, {Tremonti}, {Kennicutt},
     {Salim}, {Bothwell}, {Calzetti}, {Dalcanton}, {Dale}, {Engelbracht}, {Funes},
     {Johnson}, {Sakai}, {Skillman}, {van Zee}, {Walter}, \& {Weisz}}]{Lee09}
   {Lee}, J.~C., {Gil de Paz}, A., {Tremonti}, C., {et~al.} 2009, \apj, 706, 599
   
   \bibitem[{{Leitherer} {et~al.}(2014){Leitherer}, {Ekstr{\"o}m}, {Meynet},
     {Schaerer}, {Agienko}, \& {Levesque}}]{Leitherer14}
   {Leitherer}, C., {Ekstr{\"o}m}, S., {Meynet}, G., {et~al.} 2014, \apjs, 212, 14
   
   \bibitem[{{Leitherer} {et~al.}(1999){Leitherer}, {Schaerer}, {Goldader},
     {Delgado}, {Robert}, {Kune}, {de Mello}, {Devost}, \&
     {Heckman}}]{Leitherer99}
   {Leitherer}, C., {Schaerer}, D., {Goldader}, J.~D., {et~al.} 1999, \apjs, 123,
     3
   
   \bibitem[{{Li} {et~al.}(2017){Li}, {Ge}, {Mao}, {Cappellari}, {Long}, {Li},
     {Emsellem}, {Dutton}, {Li}, {Bundy}, {Thomas}, {Drory}, \& {Lopes}}]{Li17}
   {Li}, H., {Ge}, J., {Mao}, S., {et~al.} 2017, \apj, 838, 77
   
   \bibitem[{{Lonoce} {et~al.}(2021){Lonoce}, {Feldmeier-Krause}, \&
     {Freedman}}]{Lonoce21}
   {Lonoce}, I., {Feldmeier-Krause}, A., \& {Freedman}, W.~L. 2021, \apj, 920, 93
   
   \bibitem[{{Lonoce} {et~al.}(2023){Lonoce}, {Freedman}, \&
     {Feldmeier-Krause}}]{Lonoce23}
   {Lonoce}, I., {Freedman}, W., \& {Feldmeier-Krause}, A. 2023, arXiv e-prints,
     arXiv:2303.00044
   
   \bibitem[{{Lu} {et~al.}(2023){Lu}, {Zhu}, {Cappellari}, {Li}, {Mao}, \&
     {Xu}}]{Lu23}
   {Lu}, S., {Zhu}, K., {Cappellari}, M., {et~al.} 2023, arXiv e-prints,
     arXiv:2309.12395
   
   \bibitem[{{Lyubenova} {et~al.}(2016){Lyubenova}, {Mart{\'{\i}}n-Navarro}, {van
     de Ven}, {Falc{\'o}n-Barroso}, {Galbany}, {Gallazzi}, {Garc{\'{\i}}a-Benito},
     {Gonz{\'a}lez Delgado}, {Husemann}, {La Barbera}, {Marino}, {Mast},
     {Mendez-Abreu}, {Peletier}, {S{\'a}nchez-Bl{\'a}zquez}, {S{\'a}nchez},
     {Trager}, {van den Bosch}, {Vazdekis}, {Walcher}, {Zhu}, {Zibetti},
     {Ziegler}, {Bland-Hawthorn}, \& {CALIFA Collaboration}}]{Lyubenova16}
   {Lyubenova}, M., {Mart{\'{\i}}n-Navarro}, I., {van de Ven}, G., {et~al.} 2016,
     \mnras, 463, 3220
   
   \bibitem[{{Maiolino} \& {Mannucci}(2019)}]{Maiolino19}
   {Maiolino}, R. \& {Mannucci}, F. 2019, \aapr, 27, 3
   
   \bibitem[{{Marks} {et~al.}(2012){Marks}, {Kroupa}, {Dabringhausen}, \&
     {Pawlowski}}]{Marks}
   {Marks}, M., {Kroupa}, P., {Dabringhausen}, J., \& {Pawlowski}, M.~S. 2012,
     \mnras, 422, 2246
   
   \bibitem[{{Mart{\'{\i}}n-Navarro}
     {et~al.}(2015{\natexlab{a}}){Mart{\'{\i}}n-Navarro}, {La Barbera},
     {Vazdekis}, {Ferr{\'e}-Mateu}, {Trujillo}, \& {Beasley}}]{MN15b}
   {Mart{\'{\i}}n-Navarro}, I., {La Barbera}, F., {Vazdekis}, A., {et~al.}
     2015{\natexlab{a}}, \mnras, 451, 1081
   
   \bibitem[{{Mart{\'\i}n-Navarro} {et~al.}(2019){Mart{\'\i}n-Navarro},
     {Lyubenova}, {van de Ven}, {Falc{\'o}n-Barroso}, {Coccato}, {Corsini},
     {Gadotti}, {Iodice}, {La Barbera}, {McDermid}, {Pinna}, {Sarzi}, {Viaene},
     {de Zeeuw}, \& {Zhu}}]{MN19}
   {Mart{\'\i}n-Navarro}, I., {Lyubenova}, M., {van de Ven}, G., {et~al.} 2019,
     \aap, 626, A124
   
   \bibitem[{{Mart{\'\i}n-Navarro} {et~al.}(2021){Mart{\'\i}n-Navarro}, {Pinna},
     {Coccato}, {Falc{\'o}n-Barroso}, {van de Ven}, {Lyubenova}, {Corsini},
     {Fahrion}, {Gadotti}, {Iodice}, {McDermid}, {Poci}, {Sarzi}, {Spriggs},
     {Viaene}, {de Zeeuw}, \& {Zhu}}]{MN21}
   {Mart{\'\i}n-Navarro}, I., {Pinna}, F., {Coccato}, L., {et~al.} 2021, \aap,
     654, A59
   
   \bibitem[{{Mart{\'{\i}}n-Navarro}
     {et~al.}(2015{\natexlab{b}}){Mart{\'{\i}}n-Navarro}, {Vazdekis}, {La
     Barbera}, {Falc{\'o}n-Barroso}, {Lyubenova}, {van de Ven}, {Ferreras},
     {S{\'a}nchez}, {Trager}, {Garc{\'{\i}}a-Benito}, {Mast}, {Mendoza},
     {S{\'a}nchez-Bl{\'a}zquez}, {Gonz{\'a}lez Delgado}, {Walcher}, \& {The CALIFA
     Team}}]{MN15c}
   {Mart{\'{\i}}n-Navarro}, I., {Vazdekis}, A., {La Barbera}, F., {et~al.}
     2015{\natexlab{b}}, \apjl, 806, L31
   
   \bibitem[{{Mayya}(1997)}]{Mayya97}
   {Mayya}, Y.~D. 1997, \apjl, 482, L149
   
   \bibitem[{{M{\'e}ndez-Abreu} {et~al.}(2019){M{\'e}ndez-Abreu}, {de
     Lorenzo-C{\'a}ceres}, {Gadotti}, {Fragkoudi}, {van de Ven},
     {Falc{\'o}n-Barroso}, {Leaman}, {P{\'e}rez}, {Querejeta},
     {S{\'a}nchez-Blazquez}, \& {Seidel}}]{jairo19}
   {M{\'e}ndez-Abreu}, J., {de Lorenzo-C{\'a}ceres}, A., {Gadotti}, D.~A.,
     {et~al.} 2019, \mnras, 482, L118
   
   \bibitem[{{Meurer} {et~al.}(2009){Meurer}, {Wong}, {Kim}, {Hanish}, {Heckman},
     {Werk}, {Bland-Hawthorn}, {Dopita}, {Zwaan}, {Koribalski}, {Seibert},
     {Thilker}, {Ferguson}, {Webster}, {Putman}, {Knezek}, {Doyle}, {Drinkwater},
     {Hoopes}, {Kilborn}, {Meyer}, {Ryan-Weber}, {Smith}, \&
     {Staveley-Smith}}]{Meurer09}
   {Meurer}, G.~R., {Wong}, O.~I., {Kim}, J.~H., {et~al.} 2009, \apj, 695, 765
   
   \bibitem[{{Miller} \& {Scalo}(1979)}]{Miller79}
   {Miller}, G.~E. \& {Scalo}, J.~M. 1979, \apjs, 41, 513
   
   \bibitem[{{Milone} {et~al.}(2011){Milone}, {Sansom}, \&
     {S{\'a}nchez-Bl{\'a}zquez}}]{Milone11}
   {Milone}, A.~D.~C., {Sansom}, A.~E., \& {S{\'a}nchez-Bl{\'a}zquez}, P. 2011,
     \mnras, 414, 1227
   
   \bibitem[{{Moustakas} {et~al.}(2010){Moustakas}, {Kennicutt}, {Tremonti},
     {Dale}, {Smith}, \& {Calzetti}}]{Moustakas10}
   {Moustakas}, J., {Kennicutt}, Robert~C., J., {Tremonti}, C.~A., {et~al.} 2010,
     \apjs, 190, 233
   
   \bibitem[{{Mu{\~n}oz-Mateos} {et~al.}(2015){Mu{\~n}oz-Mateos}, {Sheth},
     {Regan}, {Kim}, {Laine}, {Erroz-Ferrer}, {Gil de Paz}, {Comeron}, {Hinz},
     {Laurikainen}, {Salo}, {Athanassoula}, {Bosma}, {Bouquin}, {Schinnerer},
     {Ho}, {Zaritsky}, {Gadotti}, {Madore}, {Holwerda}, {Men{\'e}ndez-Delmestre},
     {Knapen}, {Meidt}, {Querejeta}, {Mizusawa}, {Seibert}, {Laine}, \&
     {Courtois}}]{S4G}
   {Mu{\~n}oz-Mateos}, J.~C., {Sheth}, K., {Regan}, M., {et~al.} 2015, \apjs, 219,
     3
   
   \bibitem[{{Myers} {et~al.}(2011){Myers}, {Krumholz}, {Klein}, \&
     {McKee}}]{Myers11}
   {Myers}, A.~T., {Krumholz}, M.~R., {Klein}, R.~I., \& {McKee}, C.~F. 2011,
     \apj, 735, 49
   
   \bibitem[{{Nanayakkara} {et~al.}(2017){Nanayakkara}, {Glazebrook}, {Kacprzak},
     {Yuan}, {Fisher}, {Tran}, {Kewley}, {Spitler}, {Alcorn}, {Cowley}, {Labbe},
     {Straatman}, \& {Tomczak}}]{Nanayakkara17}
   {Nanayakkara}, T., {Glazebrook}, K., {Kacprzak}, G.~G., {et~al.} 2017, \mnras,
     468, 3071
   
   \bibitem[{{Neumann} {et~al.}(2020){Neumann}, {Fragkoudi}, {P{\'e}rez},
     {Gadotti}, {Falc{\'o}n-Barroso}, {S{\'a}nchez-Bl{\'a}zquez}, {Bittner},
     {Husemann}, {G{\'o}mez}, {Grand}, {Donohoe-Keyes}, {Kim}, {de
     Lorenzo-C{\'a}ceres}, {Martig}, {M{\'e}ndez-Abreu}, {Pakmor}, {Seidel}, \&
     {van de Ven}}]{Justus2020}
   {Neumann}, J., {Fragkoudi}, F., {P{\'e}rez}, I., {et~al.} 2020, \aap, 637, A56
   
   \bibitem[{{Neumann} {et~al.}(2019){Neumann}, {Gadotti}, {Wisotzki}, {Husemann},
     {Busch}, {Combes}, {Croom}, {Davis}, {Gaspari}, {Krumpe}, {P{\'e}rez-Torres},
     {Scharw{\"a}chter}, {Smirnova-Pinchukova}, {Tremblay}, \&
     {Urrutia}}]{Neumann19}
   {Neumann}, J., {Gadotti}, D.~A., {Wisotzki}, L., {et~al.} 2019, \aap, 627, A26
   
   \bibitem[{{Ocvirk} {et~al.}(2006){Ocvirk}, {Pichon}, {Lan{\c c}on}, \&
     {Thi{\'e}baut}}]{Ocvirk06}
   {Ocvirk}, P., {Pichon}, C., {Lan{\c c}on}, A., \& {Thi{\'e}baut}, E. 2006,
     \mnras, 365, 74
   
   \bibitem[{{Parikh} {et~al.}(2018){Parikh}, {Thomas}, {Maraston}, {Westfall},
     {Goddard}, {Lian}, {Meneses-Goytia}, {Jones}, {Vaughan}, {Andrews},
     {Bershady}, {Bizyaev}, {Brinkmann}, {Brownstein}, {Bundy}, {Drory},
     {Emsellem}, {Law}, {Newman}, {Roman-Lopes}, {Wake}, {Yan}, \&
     {Zheng}}]{Parikh}
   {Parikh}, T., {Thomas}, D., {Maraston}, C., {et~al.} 2018, \mnras, 477, 3954
   
   \bibitem[{{Percival} {et~al.}(2009){Percival}, {Salaris}, {Cassisi}, \&
     {Pietrinferni}}]{Percival09}
   {Percival}, S.~M., {Salaris}, M., {Cassisi}, S., \& {Pietrinferni}, A. 2009,
     \apj, 690, 427
   
   \bibitem[{{Pessa} {et~al.}(2023){Pessa}, {Schinnerer}, {Sanchez-Blazquez},
     {Belfiore}, {Groves}, {Emsellem}, {Neumann}, {Leroy}, {Bigiel}, {Chevance},
     {Dale}, {Glover}, {Grasha}, {Klessen}, {Kreckel}, {Kruijssen}, {Pinna},
     {Querejeta}, {Rosolowsky}, \& {Williams}}]{Pessa23}
   {Pessa}, I., {Schinnerer}, E., {Sanchez-Blazquez}, P., {et~al.} 2023, arXiv
     e-prints, arXiv:2303.13676
   
   \bibitem[{{Pietrinferni} {et~al.}(2004){Pietrinferni}, {Cassisi}, {Salaris}, \&
     {Castelli}}]{basti1}
   {Pietrinferni}, A., {Cassisi}, S., {Salaris}, M., \& {Castelli}, F. 2004, \apj,
     612, 168
   
   \bibitem[{{Pietrinferni} {et~al.}(2006){Pietrinferni}, {Cassisi}, {Salaris}, \&
     {Castelli}}]{basti2}
   {Pietrinferni}, A., {Cassisi}, S., {Salaris}, M., \& {Castelli}, F. 2006, \apj,
     642, 797
   
   \bibitem[{{Piotto} \& {Zoccali}(1999)}]{Piotto99}
   {Piotto}, G. \& {Zoccali}, M. 1999, \aap, 345, 485
   
   \bibitem[{{Posacki} {et~al.}(2015){Posacki}, {Cappellari}, {Treu},
     {Pellegrini}, \& {Ciotti}}]{Posacki15}
   {Posacki}, S., {Cappellari}, M., {Treu}, T., {Pellegrini}, S., \& {Ciotti}, L.
     2015, \mnras, 446, 493
   
   \bibitem[{{Romano} {et~al.}(2017){Romano}, {Matteucci}, {Zhang},
     {Papadopoulos}, \& {Ivison}}]{Romano17}
   {Romano}, D., {Matteucci}, F., {Zhang}, Z.~Y., {Papadopoulos}, P.~P., \&
     {Ivison}, R.~J. 2017, \mnras, 470, 401
   
   \bibitem[{{Rosani} {et~al.}(2018){Rosani}, {Pasquali}, {La Barbera},
     {Ferreras}, \& {Vazdekis}}]{Rosani18}
   {Rosani}, G., {Pasquali}, A., {La Barbera}, F., {Ferreras}, I., \& {Vazdekis},
     A. 2018, \mnras, 476, 5233
   
   \bibitem[{{Salpeter}(1955)}]{Salp:55}
   {Salpeter}, E.~E. 1955, \apj, 121, 161
   
   \bibitem[{{S{\'a}nchez} {et~al.}(2017){S{\'a}nchez}, {Barrera-Ballesteros},
     {S{\'a}nchez-Menguiano}, {Walcher}, {Marino}, {Galbany}, {Bland-Hawthorn},
     {Cano-D{\'\i}az}, {Garc{\'\i}a-Benito}, {L{\'o}pez-Cob{\'a}}, {Zibetti},
     {Vilchez}, {Igl{\'e}sias-P{\'a}ramo}, {Kehrig}, {L{\'o}pez S{\'a}nchez},
     {Duarte Puertas}, \& {Ziegler}}]{Sebastianmet}
   {S{\'a}nchez}, S.~F., {Barrera-Ballesteros}, J.~K., {S{\'a}nchez-Menguiano},
     L., {et~al.} 2017, \mnras, 469, 2121
   
   \bibitem[{{S{\'a}nchez-Bl{\'a}zquez} {et~al.}(2011){S{\'a}nchez-Bl{\'a}zquez},
     {Ocvirk}, {Gibson}, {P{\'e}rez}, \& {Peletier}}]{Pat11}
   {S{\'a}nchez-Bl{\'a}zquez}, P., {Ocvirk}, P., {Gibson}, B.~K., {P{\'e}rez}, I.,
     \& {Peletier}, R.~F. 2011, \mnras, 415, 709
   
   \bibitem[{{S{\'a}nchez-Bl{\'a}zquez} {et~al.}(2006){S{\'a}nchez-Bl{\'a}zquez},
     {Peletier}, {Jim{\'e}nez-Vicente}, {Cardiel}, {Cenarro},
     {Falc{\'o}n-Barroso}, {Gorgas}, {Selam}, \& {Vazdekis}}]{Pat06}
   {S{\'a}nchez-Bl{\'a}zquez}, P., {Peletier}, R.~F., {Jim{\'e}nez-Vicente}, J.,
     {et~al.} 2006, \mnras, 371, 703
   
   \bibitem[{{Sarzi} {et~al.}(2018){Sarzi}, {Spiniello}, {La Barbera},
     {Krajnovi{\'c}}, \& {van den Bosch}}]{Sarzi18}
   {Sarzi}, M., {Spiniello}, C., {La Barbera}, F., {Krajnovi{\'c}}, D., \& {van
     den Bosch}, R. 2018, \mnras, 478, 4084
   
   \bibitem[{{Scalo}(1998)}]{Scalo98}
   {Scalo}, J. 1998, in Astronomical Society of the Pacific Conference Series,
     Vol. 142, The Stellar Initial Mass Function (38th Herstmonceux Conference),
     ed. G.~{Gilmore} \& D.~{Howell}, 201
   
   \bibitem[{{Schwarzschild}(1979)}]{Schwarzschild79}
   {Schwarzschild}, M. 1979, \apj, 232, 236
   
   \bibitem[{{Seidel} {et~al.}(2015){Seidel}, {Cacho}, {Ruiz-Lara},
     {Falc{\'o}n-Barroso}, {P{\'e}rez}, {S{\'a}nchez-Bl{\'a}zquez}, {Vogt},
     {Ness}, {Freeman}, \& {Aniyan}}]{Seidel15}
   {Seidel}, M.~K., {Cacho}, R., {Ruiz-Lara}, T., {et~al.} 2015, \mnras, 446, 2837
   
   \bibitem[{{Serra} \& {Trager}(2007)}]{Serra07}
   {Serra}, P. \& {Trager}, S.~C. 2007, \mnras, 374, 769
   
   \bibitem[{{Sharda} \& {Krumholz}(2022)}]{Sharda22}
   {Sharda}, P. \& {Krumholz}, M.~R. 2022, \mnras, 509, 1959
   
   \bibitem[{{Sliwa} {et~al.}(2017){Sliwa}, {Wilson}, {Aalto}, \&
     {Privon}}]{Sliwa17}
   {Sliwa}, K., {Wilson}, C.~D., {Aalto}, S., \& {Privon}, G.~C. 2017, \apjl, 840,
     L11
   
   \bibitem[{{Smith}(2014)}]{smith}
   {Smith}, R.~J. 2014, \mnras, 443, L69
   
   \bibitem[{{Smith}(2020)}]{Smith20}
   {Smith}, R.~J. 2020, \araa, 58, 577
   
   \bibitem[{{Smith} \& {Lucey}(2013)}]{Smith13}
   {Smith}, R.~J. \& {Lucey}, J.~R. 2013, \mnras, 434, 1964
   
   \bibitem[{{Smith} {et~al.}(2012){Smith}, {Lucey}, \& {Carter}}]{Smith12}
   {Smith}, R.~J., {Lucey}, J.~R., \& {Carter}, D. 2012, \mnras, 426, 2994
   
   \bibitem[{{Sonnenfeld} {et~al.}(2019){Sonnenfeld}, {Jaelani}, {Chan}, {More},
     {Suyu}, {Wong}, {Oguri}, \& {Lee}}]{Sonnenfeld19}
   {Sonnenfeld}, A., {Jaelani}, A.~T., {Chan}, J., {et~al.} 2019, \aap, 630, A71
   
   \bibitem[{{Spiniello} {et~al.}(2014){Spiniello}, {Trager}, {Koopmans}, \&
     {Conroy}}]{Spiniello2013}
   {Spiniello}, C., {Trager}, S., {Koopmans}, L.~V.~E., \& {Conroy}, C. 2014,
     \mnras, 438, 1483
   
   \bibitem[{{Spiniello} {et~al.}(2015){Spiniello}, {Trager}, \&
     {Koopmans}}]{Spiniello15}
   {Spiniello}, C., {Trager}, S.~C., \& {Koopmans}, L.~V.~E. 2015, \apj, 803, 87
   
   \bibitem[{{Spiniello} {et~al.}(2012){Spiniello}, {Trager}, {Koopmans}, \&
     {Chen}}]{spiniello12}
   {Spiniello}, C., {Trager}, S.~C., {Koopmans}, L.~V.~E., \& {Chen}, Y.~P. 2012,
     \apjl, 753, L32
   
   \bibitem[{{Tang} \& {Worthey}(2017)}]{Tang17}
   {Tang}, B. \& {Worthey}, G. 2017, \mnras, 467, 674
   
   \bibitem[{{Tanvir} {et~al.}(2022){Tanvir}, {Krumholz}, \&
     {Federrath}}]{Tanvir22}
   {Tanvir}, T.~S., {Krumholz}, M.~R., \& {Federrath}, C. 2022, arXiv e-prints,
     arXiv:2206.04999
   
   \bibitem[{{Thomas} {et~al.}(2005){Thomas}, {Maraston}, {Bender}, \& {Mendes de
     Oliveira}}]{Thomas05}
   {Thomas}, D., {Maraston}, C., {Bender}, R., \& {Mendes de Oliveira}, C. 2005,
     \apj, 621, 673
   
   \bibitem[{{Thomas} {et~al.}(2011){Thomas}, {Saglia}, {Bender}, {Thomas},
     {Gebhardt}, {Magorrian}, {Corsini}, {Wegner}, \& {Seitz}}]{thomas11}
   {Thomas}, J., {Saglia}, R.~P., {Bender}, R., {et~al.} 2011, \mnras, 415, 545
   
   \bibitem[{{Tinsley} \& {Gunn}(1976)}]{Tinsley76}
   {Tinsley}, B.~M. \& {Gunn}, J.~E. 1976, \apj, 203, 52
   
   \bibitem[{{Tojeiro} {et~al.}(2007){Tojeiro}, {Heavens}, {Jimenez}, \&
     {Panter}}]{Vespa}
   {Tojeiro}, R., {Heavens}, A.~F., {Jimenez}, R., \& {Panter}, B. 2007, \mnras,
     381, 1252
   
   \bibitem[{{Tortora} {et~al.}(2014){Tortora}, {La Barbera}, {Napolitano},
     {Romanowsky}, {Ferreras}, \& {de Carvalho}}]{Tortora14}
   {Tortora}, C., {La Barbera}, F., {Napolitano}, N.~R., {et~al.} 2014, \mnras,
     445, 115
   
   \bibitem[{{Tortora} {et~al.}(2013){Tortora}, {Romanowsky}, \&
     {Napolitano}}]{Tortora13b}
   {Tortora}, C., {Romanowsky}, A.~J., \& {Napolitano}, N.~R. 2013, \apj, 765, 8
   
   \bibitem[{{Trager} {et~al.}(2000){Trager}, {Faber}, {Worthey}, \&
     {Gonz{\'a}lez}}]{Trager00}
   {Trager}, S.~C., {Faber}, S.~M., {Worthey}, G., \& {Gonz{\'a}lez}, J.~J. 2000,
     \aj, 119, 1645
   
   \bibitem[{{Treu} {et~al.}(2010){Treu}, {Auger}, {Koopmans}, {Gavazzi},
     {Marshall}, \& {Bolton}}]{Treu}
   {Treu}, T., {Auger}, M.~W., {Koopmans}, L.~V.~E., {et~al.} 2010, \apj, 709,
     1195
   
   \bibitem[{{van den Bosch} {et~al.}(2008){van den Bosch}, {van de Ven},
     {Verolme}, {Cappellari}, \& {de Zeeuw}}]{Remco08}
   {van den Bosch}, R.~C.~E., {van de Ven}, G., {Verolme}, E.~K., {Cappellari},
     M., \& {de Zeeuw}, P.~T. 2008, \mnras, 385, 647
   
   \bibitem[{{van Dokkum} {et~al.}(2017){van Dokkum}, {Conroy}, {Villaume},
     {Brodie}, \& {Romanowsky}}]{vdk17}
   {van Dokkum}, P., {Conroy}, C., {Villaume}, A., {Brodie}, J., \& {Romanowsky},
     A.~J. 2017, \apj, 841, 68
   
   \bibitem[{{van Dokkum} \& {Conroy}(2010)}]{vandokkum}
   {van Dokkum}, P.~G. \& {Conroy}, C. 2010, \nat, 468, 940
   
   \bibitem[{{Vazdekis} {et~al.}(1996){Vazdekis}, {Casuso}, {Peletier}, \&
     {Beckman}}]{vazdekis96}
   {Vazdekis}, A., {Casuso}, E., {Peletier}, R.~F., \& {Beckman}, J.~E. 1996,
     \apjs, 106, 307
   
   \bibitem[{{Vazdekis} {et~al.}(2015){Vazdekis}, {Coelho}, {Cassisi},
     {Ricciardelli}, {Falc{\'o}n-Barroso}, {S{\'a}nchez-Bl{\'a}zquez}, {La
     Barbera}, {Beasley}, \& {Pietrinferni}}]{Vazdekis15}
   {Vazdekis}, A., {Coelho}, P., {Cassisi}, S., {et~al.} 2015, \mnras, 449, 1177
   
   \bibitem[{{Vazdekis} {et~al.}(1997){Vazdekis}, {Peletier}, {Beckman}, \&
     {Casuso}}]{vazdekis:97}
   {Vazdekis}, A., {Peletier}, R.~F., {Beckman}, J.~E., \& {Casuso}, E. 1997,
     \apjs, 111, 203
   
   \bibitem[{{Vazdekis} {et~al.}(2010){Vazdekis}, {S{\'a}nchez-Bl{\'a}zquez},
     {Falc{\'o}n-Barroso}, {Cenarro}, {Beasley}, {Cardiel}, {Gorgas}, \&
     {Peletier}}]{miles}
   {Vazdekis}, A., {S{\'a}nchez-Bl{\'a}zquez}, P., {Falc{\'o}n-Barroso}, J.,
     {et~al.} 2010, \mnras, 404, 1639
   
   \bibitem[{{Walcher} {et~al.}(2009){Walcher}, {Coelho}, {Gallazzi}, \&
     {Charlot}}]{Walcher09}
   {Walcher}, C.~J., {Coelho}, P., {Gallazzi}, A., \& {Charlot}, S. 2009, \mnras,
     398, L44
   
   \bibitem[{{Walcher} {et~al.}(2015){Walcher}, {Coelho}, {Gallazzi}, {Bruzual},
     {Charlot}, \& {Chiappini}}]{Walcher15}
   {Walcher}, C.~J., {Coelho}, P.~R.~T., {Gallazzi}, A., {et~al.} 2015, \aap, 582,
     A46
   
   \bibitem[{{Wegner} {et~al.}(2012){Wegner}, {Corsini}, {Thomas}, {Saglia},
     {Bender}, \& {Pu}}]{wegner12}
   {Wegner}, G.~A., {Corsini}, E.~M., {Thomas}, J., {et~al.} 2012, \aj, 144, 78
   
   \bibitem[{{Wilkinson} {et~al.}(2017){Wilkinson}, {Maraston}, {Goddard},
     {Thomas}, \& {Parikh}}]{Wilkinson17}
   {Wilkinson}, D.~M., {Maraston}, C., {Goddard}, D., {Thomas}, D., \& {Parikh},
     T. 2017, \mnras, 472, 4297
   
   \bibitem[{{Worthey} {et~al.}(1994){Worthey}, {Faber}, {Gonzalez}, \&
     {Burstein}}]{Worthey94b}
   {Worthey}, G., {Faber}, S.~M., {Gonzalez}, J.~J., \& {Burstein}, D. 1994,
     \apjs, 94, 687
   
   \bibitem[{{Zhang} {et~al.}(2018){Zhang}, {Romano}, {Ivison}, {Papadopoulos}, \&
     {Matteucci}}]{Zhang18}
   {Zhang}, Z.-Y., {Romano}, D., {Ivison}, R.~J., {Papadopoulos}, P.~P., \&
     {Matteucci}, F. 2018, \nat, 558, 260
   
   \bibitem[{{Zhou} {et~al.}(2019){Zhou}, {Mo}, {Li}, {Zheng}, {Li}, {Du}, {Mao},
     {Parikh}, {Lane}, \& {Thomas}}]{Zhou19}
   {Zhou}, S., {Mo}, H.~J., {Li}, C., {et~al.} 2019, \mnras, 485, 5256
   
   \bibitem[{{Zhu} {et~al.}(2018){Zhu}, {van de Ven}, {van den Bosch}, {Rix},
     {Lyubenova}, {Falc{\'o}n-Barroso}, {Martig}, {Mao}, {Xu}, {Jin}, {Obreja},
     {Grand}, {Dutton}, {Macci{\`o}}, {G{\'o}mez}, {Walcher},
     {Garc{\'\i}a-Benito}, {Zibetti}, \& {S{\'a}nchez}}]{Ling18b}
   {Zhu}, L., {van de Ven}, G., {van den Bosch}, R., {et~al.} 2018, Nature
     Astronomy, 2, 233
   
   \bibitem[{{Zhuang} {et~al.}(2019){Zhuang}, {Leaman}, {van de Ven}, {Zibetti},
     {Gallazzi}, {Zhu}, {Falc{\'o}n-Barroso}, \& {Lyubenova}}]{Yulong}
   {Zhuang}, Y., {Leaman}, R., {van de Ven}, G., {et~al.} 2019, \mnras, 483, 1862
   
   \bibitem[{{Zoccali} {et~al.}(2000){Zoccali}, {Cassisi}, {Frogel}, {Gould},
     {Ortolani}, {Renzini}, {Rich}, \& {Stephens}}]{Zoccali00}
   {Zoccali}, M., {Cassisi}, S., {Frogel}, J.~A., {et~al.} 2000, \apj, 530, 418
   
   \end{thebibliography}

\begin{appendix}

\section{Fitting uncertainties and systematics trends} \label{sec:system}

\begin{figure}[!h]
   \centering
   \includegraphics[width=9cm]{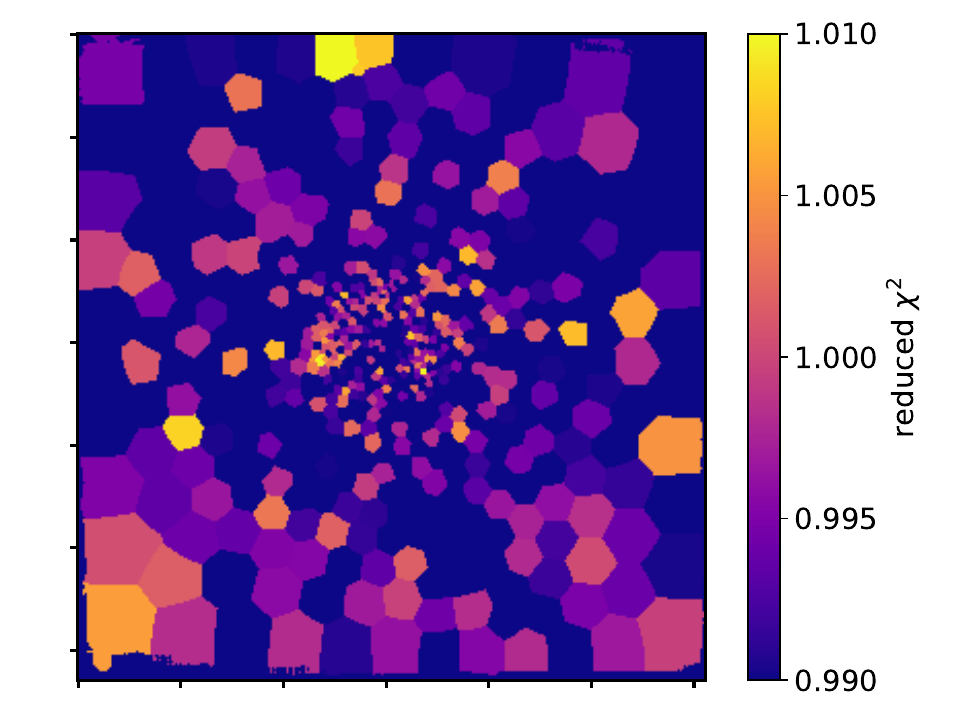}
   \caption{\textbf{Reduced $\chi^2$ map for NGC\,3351. No obvious features are visible in the map, suggesting that our fitting approach is capturing reasonably well the expected uncertainties in the stellar population properties.}}
   \label{fig:chi}
\end{figure}

\textbf{The wide range of stellar population properties (ages, star formation histories, metallicities etc.) exhibited by the stellar populations of NGC\,3351 pose a challenge for our modelling approach. Figure~\ref{fig:chi} shows the reduced $\chi^2$ resulting from our best fitting solutions. No clear structures are evident from this figure, suggesting that the uncertainties in the recovered stellar population parameters are reasonably well approximated. This is mostly due to the fact that our best-fitting model allows for a re-scaling of the assumed error spectrum (see details in S~\ref{sec:fit}). The reduced $\chi^2$ is rather homogeneous and around the expected $\chi^2\eqsim1$ value, although this does not take into account systematic uncertainties. }

\begin{figure}[!h]
   \centering
   \includegraphics[width=9cm]{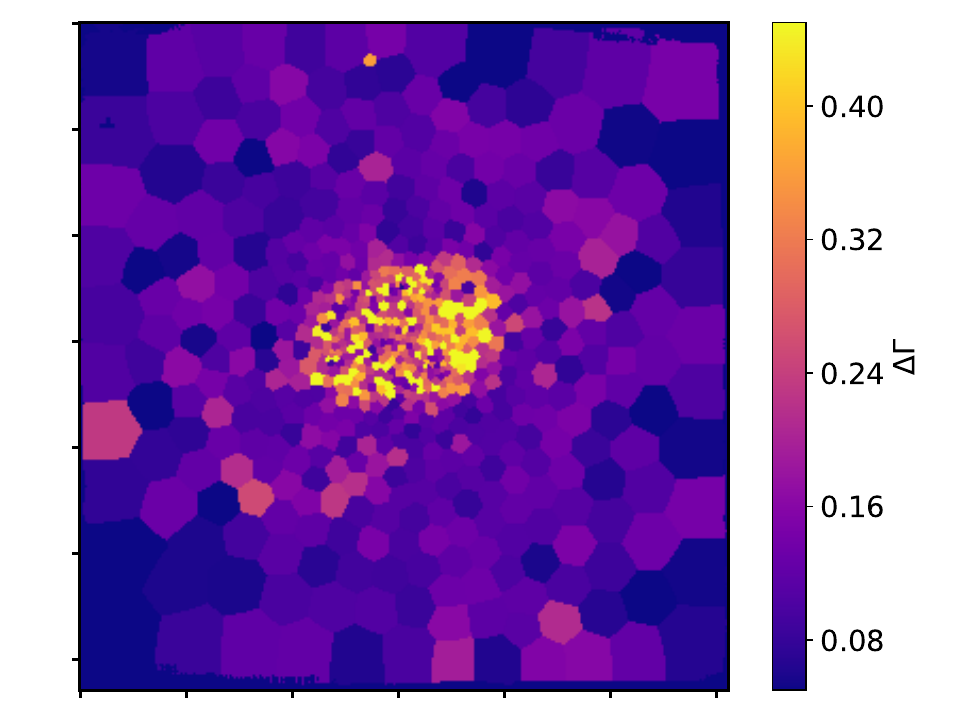}
   \caption{\textbf{Measured 1$\sigma$ uncertainty in the best-fitting IMF slope values. IMF slope values are well constrained across the whole MUSE field of view, particularly outside the star forming ring, where old stellar populations dominate the integrated flux. In the center of NGC\,3351 uncertainties are significantly larger due to the more extended star formation histories.}}
   \label{fig:imf_err}
\end{figure}

\textbf{In addition, Fig.~\ref{fig:imf_err} contains the estimated 1$\sigma$ uncertainty in the best-fitting IMF slope value. Across the relatively old populations of the disk, the errors in the measured $\Gamma$ values are relatively small. For the central regions of NGC\,3351, the young and complex star formation histories of the stellar populations prevents more accurate IMF measurements. It is worth highlighting however that our fitting scheme is able to capture this increment in the intrinsic degeneracies of the inversion problem when dealing with more extended star formation histories.}

\section{Incomplete modeling approach} \label{sec:wrong}

In \S~\ref{sec:fit} we describe our fitting approach, in particular the improvements made over the standard SSP assumption that is behind IMF measurements of quiescent ETGs based on their integrated spectra. Two key ingredients are necessary to successfully measure the IMF slope of younger populations with extended star formation histories. First, it is mandatory to extend the models towards younger populations. These very young populations are not present in quiescent systems but can have a dramatic impact on the observed spectra (see e.g. Fig.~\ref{fig:indices}). Even if the modeling of these populations is challenging and still far from being as well-understood as older stellar populations, its inclusion in the fitting scheme is required. In addition, it is also necessary to allow for an extended star formation history to fit the observed data \citep[see e.g. Figure 25 in][]{Vazdekis15}. 

\begin{figure}[!h]
      \centering
      \includegraphics[width=9cm]{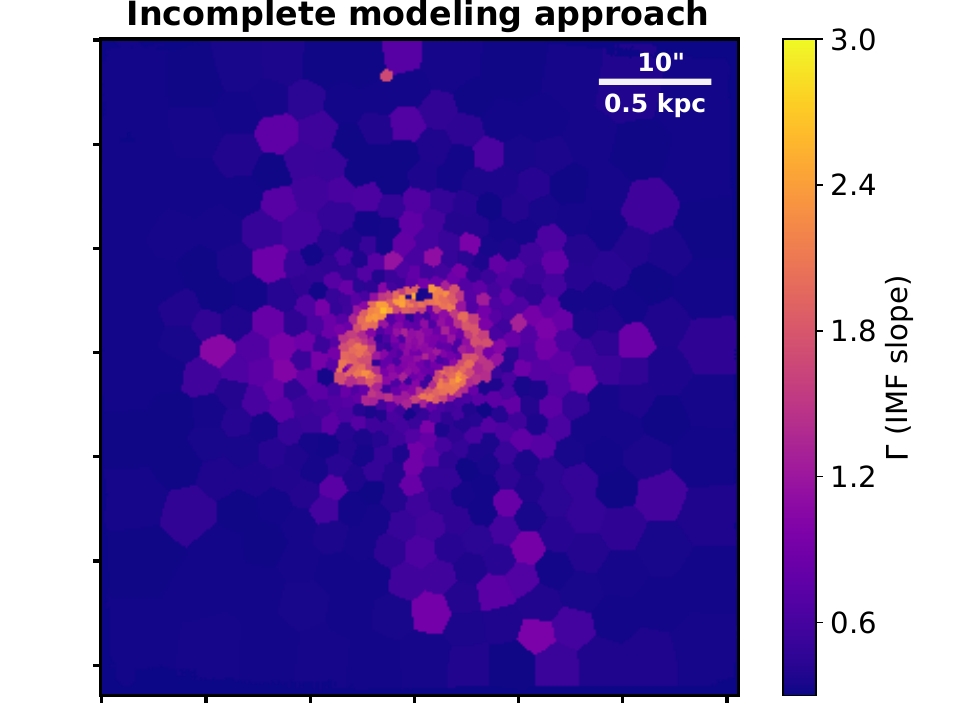}
      \caption{Inaccurate IMF slope map of NGC\,3351, obtained assuming that its stellar populations are well-characterized by an SSP and only including stars older than 63 Myr. This approach leads to a biased IMF measurement in the central regions of the galaxy where young populations with extended star formation histories dominate the optical spectrum.}
      \label{fig:IMF_wrong}
   \end{figure}

In order to exemplify the importance of these improvements, Fig.~\ref{fig:IMF_wrong} shows the recovered IMF map of NGC\,3351 following a SSP-based fitting approach that does not include populations younger than 63 Myr. Clearly visible in the IMF map, the star-forming ring of the galaxy seems to have a steep IMF slope, similar to what is observed in the centers of massive ETGs. However, this is purely a systematic bias due to the limitations of the modeling approach. A combination of the effect shown in Fig.~\ref{fig:offgrid}, where the SSP approach can lead to steeper IMF slopes in composite stellar populations, plus the incomplete sampling of the supergiant phase, push the recovered IMF slopes to unrealistic values. 

\section{Alternative IMF functional forms} \label{sec:bimodal}

\begin{figure}[!h]
   \centering
   \includegraphics[width=9cm]{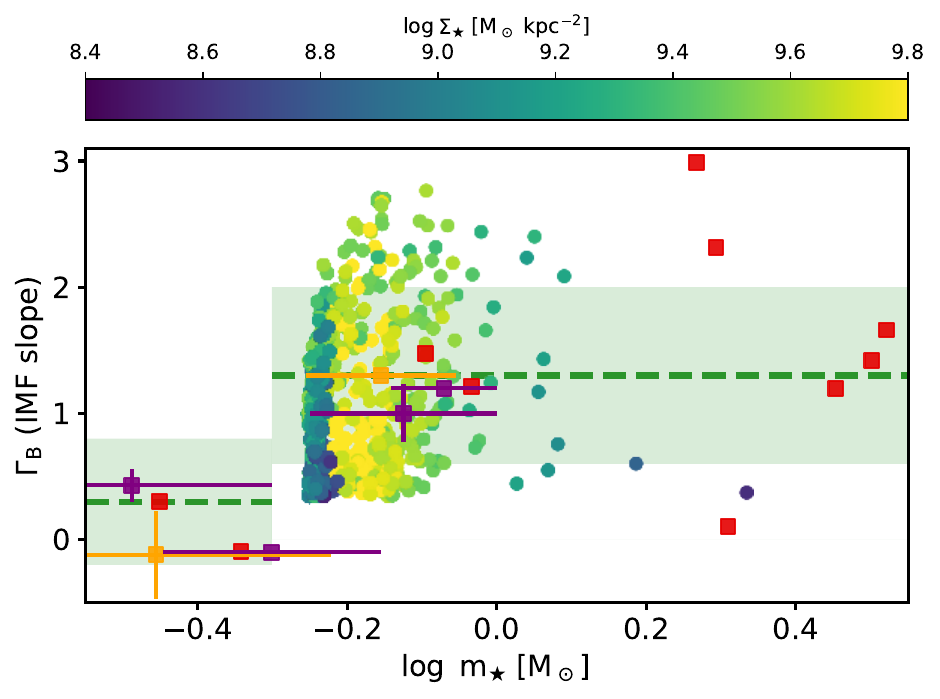}
   \caption{The measured slope of the IMF assuming a broken power law parametrization ($\Gamma_\mathrm{B}$) is shown as a function of the (logarithmic) stellar mass. This figure is the same as Fig.~\ref{fig:shape} but under a different IMF functional form. The overall trends discussed for Fig.~\ref{fig:shape} are preserved, demonstrating that the obtained results are robust against the adopted IMF shape.}
   \label{fig:bimodal}
\end{figure}

\textbf{For our stellar population analysis we have adopted a single-power law IMF parametrization. Such a simple function form has clear advantages as describe in the sections above but it certainly not realistic. Within the MILES stellar population synthesis models, there is an additional IMF parametrization, the so-called bi-modal IMF. This IMF, as defined in \citet{vazdekis96}, is characterized by the slope of the high-mass end $\Gamma_\mathrm{B}$ and it tapers out for stellar masses below $\sim0.5$\msun. When  $\Gamma_\mathrm{B} = 1.3$, the bi-modal parametrization essentially corresponds to a \citet{kroupa} IMF. For completeness, we repeated our stellar population analysis using the a broken power law functional form. Fig.~\ref{fig:bimodal} shows the relation between this bi-modal IMF slope $\Gamma_\mathrm{B}$ and the mean stellar mass of the different Voronoi bins, replicating Fig.~\ref{fig:shape} in the main text but under the alternative IMF parametrization. }

The trends shown in Fig.~\ref{fig:bimodal} are essentially the same as in Fig.~\ref{fig:shape}, with the only difference that the recovered $\Gamma_\mathrm{B}$ values expand over a slightly larger dynamical range. This is a well-known difference between the two parametrizations \citep[see e.g.][]{ferreras} and does not affect the overall interpretation of the maps. In general, the measured $\Gamma_\mathrm{B}$ values are around the Milky Way expectations, with regions of relatively lower densities (and metallicities) showing on average flatter IMF slopes. 

\end{appendix}

\end{document}